\newcommand{\PaperTitle}{Political Leanings in Web3 Betting: Decoding the Interplay of Political and Profitable Motives}

\documentclass[10pt,sigconf,screen,letterpaper,nonacm]{acmart}

\usepackage{amsmath}%
\usepackage{graphicx}%
\usepackage{multirow}%
\usepackage{amsfonts}
\usepackage{amsthm}%
\usepackage[title]{appendix}%
\usepackage{xcolor}%
\usepackage{textcomp}%
\usepackage{manyfoot}%
\usepackage{booktabs}%
\usepackage{algorithm}%
\usepackage{algorithmicx}%
\usepackage{algpseudocode}%
\usepackage{listings}%
\usepackage{array}
\usepackage{subcaption}
\usepackage{booktabs}
\usepackage{placeins}

\AtBeginDocument{%
  }

\setcopyright{acmlicensed}
\acmDOI{XXXXXXX.XXXXXXX}

\acmConference[Conference acronym 'XX]{Make sure to enter the correct
  conference title from your rights confirmation emai}{June 03--05,
  2018}{Woodstock, NY}
\acmISBN{978-1-4503-XXXX-X/18/06}

\begin{document}

\title[Political Leanings in Web3 Betting]{\PaperTitle}

\author{Hongzhou Chen}
\email{hongzhouchen1@link.cuhk.edu.cn}
\orcid{0000-0002-8310-3202}
\affiliation{%
  \institution{The Chinese University of Hong Kong, Shenzhen; Mohamed bin Zayed University of Artificial Intelligence}
  \country{China}
}

\author{Xiaolin Duan}
\email{duanxiaolin@cuhk.edu.cn}
\orcid{0000-0002-7284-9586}
\affiliation{%
  \institution{The Chinese University of Hong Kong, Shenzhen}
  \country{China}
  }

\author{Abdulmotaleb El Saddik}
\email{elsaddik@uottawa.ca}
\orcid{0000-0002-7690-8547}
\affiliation{%
  \institution{Mohamed bin Zayed University of Artificial Intelligence; University of Ottawa}
  \country{United Arab Emirates}
}

\author{Wei Cai}
 \email{weicai@ieee.org}
\orcid{0000-0002-4658-0034}
\affiliation{%
 \institution{The Chinese University of Hong Kong, Shenzhen}
 \country{China}
 }

 \authornote{Wei Cai is the corresponding author.}

\renewcommand{\shortauthors}{Chen et al.}

\begin{abstract}
Harnessing the transparent blockchain user behavior data, we construct the Political Betting Leaning Score (PBLS) to measure political leanings based on betting within Web3 prediction markets. Focusing on Polymarket and starting from the 2024 U.S. Presidential Election, we synthesize behaviors over 15,000 addresses across 4,500 events and 8,500 markets, capturing the intensity and direction of their political leanings by the PBLS. We validate the PBLS through internal consistency checks and external comparisons. We uncover relationships between our PBLS and betting behaviors through over 800 features capturing various behavioral aspects. A case study of the 2022 U.S. Senate election further demonstrates the ability of our measurement while decoding the dynamic interaction between political and profitable motives. Our findings contribute to understanding decision-making in decentralized markets, enhancing the analysis of behaviors within Web3 prediction environments. The insights of this study reveal the potential of blockchain in enabling innovative, multidisciplinary studies and could inform the development of more effective online prediction markets, improve the accuracy of forecast, and help the design and optimization of platform mechanisms. The data and code for the paper are accessible at the following link: https://github.com/anonymous.
\end{abstract}

\keywords{Political leaning, Profit motive, Web3 prediction markets, Betting behavior, Polymarket}

\maketitle

\section{Introduction}\label{intro}
The rapid development of blockchain networks has given rise to a new form of Web3 prediction markets, which utilize the principles of decentralized finance (DeFi) and the wisdom of crowds to facilitate the aggregation of diverse opinions and predictions~\cite{chen2023web3,buckley2022blockchain}. Compared with traditional online prediction markets, like Iowa Electronic Markets and Betfair, Web3 platforms utilize blockchain's transparency, immutability, and anonymity, facilitate participation, diminish manipulation risks, and ensure verifiable online records~\cite{chen2023web3}. This supports the creation of detailed, aggregated, and privacy-preserving datasets about betting behaviors. Moreover, studying Web3 bettings has the potential to revolutionize how we study and understand political leaning, which profoundly shapes individuals' attitudes and decisions~\cite{weatherford1982interpersonal,wong2016quantifying}, particularly in the interplay of political and profitable motives.

Our \textbf{research question} is: \textit{Can individuals' political leanings be decoded from their betting behaviors in Web3 prediction markets, and if so, how does the interplay of political and profitable motives influence participants' betting strategies and market outcomes?} To answer it, we focus on Polymarket, the largest Web3 prediction betting platform~\cite{buckley2022blockchain} to examine users' betting patterns across various events and markets.

By leveraging the rich, granular data, we construct the Political Betting Leaning Score (PBLS), to quantify user addresses' political leanings based on their on-chain betting activities across various markets. Our approach draws upon the theoretical foundations of rational choice theory, which posits that individuals' decisions aimed at maximizing personal utility encapsulate their political preferences~\cite{becker1976economic}. This diverges from traditional political leaning assessments by prioritizing the ``revealed preferences'' concept, suggesting that financial investments serve as a more reliable indicator of political inclination~\cite{samuelson1938note}. In designing the PBLS, we incorporate multiple dimensions of user activity, such as the direction and amount of bets, the timing of trades relative to market outcomes, and the consistency of betting patterns across different political events. By aggregating these behavioral signals, the PBLS aims to provide a robust and comprehensive measure of a user's political leaning. 

Our study has four main steps: (1) We explore the overview and accuracy of political prediction on Polymarket, providing study context. (2) We construct and compute the PBLS for participants based on the 2024 U.S. Presidential Election event. We validate the PBLS through internal consistency checks using the popular vote event and external comparisons with real-world polls. (3) We conduct comprehensive feature engineering, constructing over 800 features that capture various dimensions of user behavior on Polymarket. Through correlation analysis, we identify significant relationships between these features and the PBLS, revealing the complex patterns in how political leanings manifest in betting activities. We then employ machine learning techniques to predict the PBLS for more than 15,000 user addresses, extending our analysis to the broader Polymarket user base. (4) By a case study of the 2022 U.S. Senate election, we demonstrate the ability of PBLS and the interplay between political leanings and profit motives, acknowledging their joint influence on betting behaviors and providing a more thorough decoding of political leanings revealed through Web3 betting.

This work contributes to Internet end-user behavioral studies. First, we introduce a novel method for inferring political leanings from economic behavior by leveraging the unique capabilities of Web3 prediction markets. Second, the PBLS captures revealed preferences rather than statements, providing a more objective and granular assessment. Third, we shed light on how political and profitable motives shape decision-making in decentralized market environments. Our work expands behavior research methods and demonstrates the potential of blockchain in enabling innovative, multidisciplinary studies. The insights of this study could inform the development of more effective prediction markets, improve the accuracy of forecast, and help the design and optimization of online platform mechanisms. The data and code for the paper are accessible at https://github.com/anonymous.

\section{Background and Related Work}\label{related_work}
\subsection{Measure Political Leanings Online}
Political leanings profoundly shape attitudes and behaviors across various domains~\cite{weatherford1982interpersonal,wong2016quantifying}. Traditional political leanings measurements, such as surveys and self-reports, often suffer from social desirability bias and lack of granularity, difficulty capturing real behaviors~\cite{grimm2010social,cowan2018could}. Digital platforms open new avenues for studying political leanings through online activity analysis. However, these methods often rely on the discourse contents, which can be noisy, ill-structured, and constrained by echo chambers~\cite{boutyline2017social,wong2013quantifying}.

Online prediction platforms, such as the Iowa Electronic Markets and Betfair, have emerged as a promising alternative to forecast and reveal collective beliefs~\cite{berg2008prediction,surowiecki2005wisdom}. These markets allow participants to bet based on the outcomes of specific events, such as elections or policy decisions, often outperforming traditional methods like polls or expert opinions. Crucially, the incentive structure of prediction markets encourages participants to reveal their true beliefs and expectations, as they have a financial stake in their prediction accuracy~\cite{arrow2012social,manski2006interpreting}. This makes prediction markets valuable for studying political leanings, as trading behavior can provide a more direct and unbiased measure of political preferences, while it is essential to consider how political and profitable motives interact to shape behaviors and outcomes. 

\subsection{Polymarket, Web3 Prediction Market}
Traditional prediction markets have limitations on centralization and lack transparency, hindering access and comprehensive analysis of granular user behaviors~\cite{yang2015information} and raising concerns about potential manipulation~\cite{abramowicz2006prediction}. Additionally, the aggregation and analysis of behavior data in traditional prediction markets, where identity verification is compulsory, can be constrained by privacy concerns and regulatory restrictions~\cite{ozimek2014regulation}. The emergence of decentralized Web3, which is based on blockchain, offers solutions to the constraints of conventional platforms. Web3 platforms, utilizing blockchain's transparency, immutability, and anonymity, facilitate participation, diminish manipulation risks, and ensure verifiable online records~\cite{chen2023web3}. This supports the creation of detailed, aggregated, and privacy-preserving datasets about user behavior on Web3 prediction markets. 
\begin{figure}[htbp]
\centering 
  \includegraphics[width=.8\linewidth]{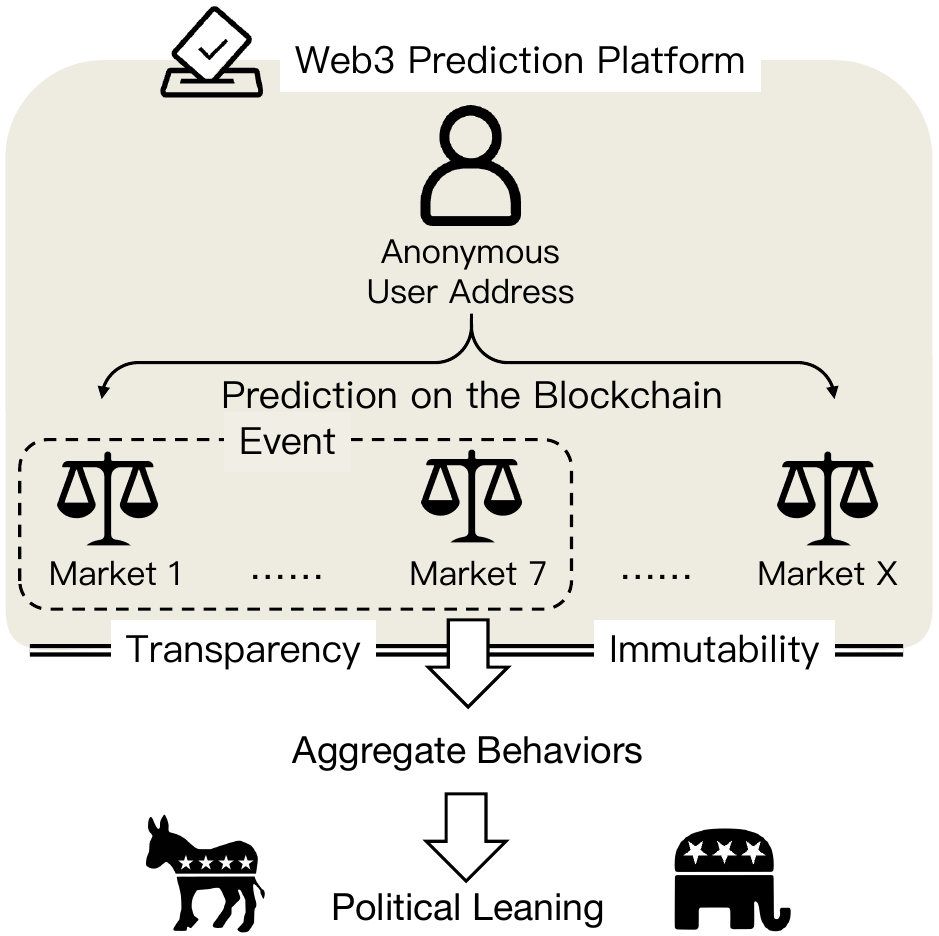} 
  \caption{Decoding political leanings through on-chain betting behaviors. The blockchain enables transparent and immutable records of anonymous behaviors across markets. By aggregating, we could construct a quantitative measure of political leaning.}
  \label{fig:background}
\end{figure}

As the largest Web3 prediction market, Polymarket~\cite{buckley2022blockchain} is built on the Polygon blockchain and leverages the principles of decentralized finance (DeFi) and the wisdom of crowds to facilitate the aggregation of diverse opinions and predictions. On Polymarket, users propose and create events or markets, where events serve as collections of related markets that focus on specific outcomes within the event. For instance, an event might revolve around the U.S. Presidential Election, while the associated markets could address more granular aspects such as the ``Yes'' or ``No'' on specific candidates. 

\subsection{Decoding Online Betting Behaviors}
Theoretically, our methodology for decoding individuals' online betting behaviors is anchored in rational choice theory, which posits that individuals' decisions aimed at maximizing personal utility, encapsulate their political preferences~\cite{becker1976economic}. This diverges from traditional political leaning assessments, such as surveys or social media analyses, by prioritizing financial factors, embodying the ``revealed preferences'' concept~\cite{samuelson1938note} as a more reliable indicator of political leaning. 

Previous studies have successfully employed financial and aggregated digital behavioral data to infer political leanings, our methodology innovates by leveraging the unique capabilities of blockchain technology and the mechanisms of prediction markets (Figure~\ref{fig:background}). For instance, Bonaparte and Kumar~\cite{bonaparte2013political} illustrate how individuals' political affiliations can influence their financial decisions in stock markets, echoing the underlying premise of our approach that economic behaviors reflect political inclinations. Similarly, Conover et al.~\cite{conover2011political} demonstrate the potential of aggregated digital footprints, such as social media interactions. Our work applies these principles to analyzing betting behavior in blockchain-based prediction markets, capitalizing on the transparent and immutable record afforded by blockchain~\cite{sillaber2017life}.

\section{Methodology}\label{method}
\subsection{Data Collection}\label{method:data}
We deployed an entities-indexed and live updated blockchain subgraph of Polymarket\footnote{https://api.studio.thegraph.com/anonymous} including more than 56M blocks and 7M entities, convenient for future research. For early transactions not captured by the subgraph, we obtained supplementary blockchain data using Polygonscan\footnote{https://polygonscan.com}. Specific event and market data (such as slug and prediction details) was retrieved using Polymarket API\footnote{https://docs.polymarket.com}. We employed a fine-tuned model to fill in the missing IDs (details in Appendix~\ref{app:fine}).

We collected Polymarket data from 4,592 ``events'' (collections of related ``markets'') and 8,742 ``markets'', spanning the first market on October 2, 2020, until February 27, 2024. Our dataset includes over 1.8 million betting records for 15,555 unique blockchain addresses participating in political betting on Polymarket. These records include transaction hashes, timestamps, names of events and markets, purchase directions (buy or sell), bet targets (e.g., Yes or No, Republican or Democratic, etc.), share sizes, prices, and amounts (in USDC). Their activities involved 4,592 (100\%) events and 8,569 (98\%) markets, indicating that political market participants engaged in various types of prediction betting. In other words, Polymarket provides extensive behavioral dimensions for measuring political leanings through betting behaviors that are even if not directly related.

Notably, as a double-edged sword, we cannot obtain demographic data about Polymarket users due to blockchain anonymity. During interpreting, we must consider that Polymarket users may not represent the general population but rather a subsample with biases. Nevertheless, the advantage of blockchain data lies in its objectivity and behavioral-level granularity, providing opportunities for measuring political leanings that other ways can hardly match.

\subsection{Accuracy of Political Betting Markets}\label{method:accuracy}
Among the 1,104 political-related markets in our dataset, we focused on the already closed markets. For each of them, we calculated the Discrete Log Score (DLS) to measure the accuracy of the predictions. As the variant of Brier score~\cite{rufibach2010use}, the DLS is widely applied to score continuous binary outcomes that resolve out-of-bounds. Given a prediction probability $P$, the DLS is computed as follows:
\begin{equation}
DLS = \begin{cases}
\log_2(P) + 1, & \text{if the event occurred,} \\
\log_2(1 - P) + 1, & \text{otherwise.}
\end{cases}
\end{equation}

\noindent
The higher DLS indicates better prediction. The maximum possible is 1, corresponding to a perfect prediction (i.e., 100\% confidence in the correct outcome). A maximally uncertain prediction of $P = 0.5$ yields a score of $S = 0$. A score greater than 0 indicates that the prediction market performed better than a maximally uncertain prediction (i.e., random guessing with 50\% confidence in each outcome) and vice versa. 

We employed an Exponential Weighted Moving Average (EWMA) of the DLS to visualize the trend in prediction accuracy over time. The EWMA is a time-decayed average, where each resolved question is weighted 5\% less than the question resolved immediately following. This approach emphasizes recent prediction performance while considering historical accuracy. We investigated the relationships between the EWMA DLS and two key characteristics: the number of participants and the total betting volumes, which were normalized to account for market size and duration differences. Linear regression analyses were performed to estimate the coefficients and the $R^2$ scores, which indicate the proportion of variance in the EWMA explained by each characteristic.

\subsection{Scoring Political Betting Leaning}\label{method:pbls}
We introduce the Political Betting Leaning Score (PBLS), a quantitative measure designed to assess the political leanings of individuals based on their betting activities within the predictions directly reflecting their political leaning, like the 2024 U.S. presidential election on Polymarket. 

The PBLS is calculated for each user address $u$ as follows:
\begin{equation}
\text{PBLS}u = \frac{\sum_{i \in T_u} w_{\text{p}, i} \cdot w_{\text{t}, i} \cdot w_{\text{amt}, i} \cdot w_{\text{pty}, i}}{\sum_{i \in T_u} |w_{\text{p}, i} \cdot w_{\text{t}, i} \cdot w_{\text{amt}, i}|} \cdot w_{\text{freq}, u}
\end{equation}
, where $T_u$ represents the set of trades executed by user address $u$. Each trade $i$ is associated with a set of weights:

\textbf{Price Weight ($w_{\text{p}, i}$)}: Market prices reflect the collective expectations regarding event outcomes~\cite{manski2006interpreting}. Scholars have demonstrated the significance of market prices in reflecting political leanings, which are directly related to confidence or leaning in outcome possibility and also form through betting behaviors~\cite{erikson2012markets,wolfers2006interpreting}. This weight is formulated as

\begin{equation}
w_{\text{p}, i} = \begin{cases}
\frac{|p_i - 0.5|}{0.5}, & \text{if} |p_i - 0.5| > \theta \\
0, & \text{otherwise}
\end{cases}
\end{equation},
where uncertain prediction (i.e., 0.5) has the 0 weight and  $\theta = 0.005$ acts as a threshold to discount trades with prices near certainty (i.e., close to 0 or 1). The $\theta$ is rooted in behavioral economics, particularly in the concept of arbitrage in prediction markets. It posits that participants tend to exploit minute price differentials in near certainties, akin to the ``free-rider'' problem~\cite{hanson2003combinatorial,schram1991people}. By setting a threshold, we aim to emphasize trades reflective of genuine political motives rather than opportunistic gains. 

\textbf{Time Decay Weight ($w_{\text{t}, i}$)}: highlights the relevance of recent activities in election forecasting~\cite{fujiwara2016habit}. Its formula is
\begin{equation}
w_{\text{t}, i} = \exp\left(-\frac{\log(2)}{h} \cdot (t_{\text{end}} - t_i)\right)
\end{equation},
where $h$ is the half-life. Essentially, this weighting aligns with the recency effect, which suggests that recent information tends to impact and reflect perceptions and decisions~\cite{murdock1962serial} substantially. In their study of online search behavior, Ding et al.~\cite{ding2009pagerank} employed a 7-day half-life. Since political leanings should have a longer relevance time scale than other online behaviors, we set the $h$ to 14 days to balance the decay effect and the persistence of political attitudes.

\textbf{Amount Weight ($w_{\text{amt}, i}$)}: The betting amount is directly related to utility, like engagement or risk preference, widely used in prediction and decision studies~\cite{rothschild2016trading}. Its formula is
\begin{equation}
w_{\text{amt}, i} = \pm \log(1 + a_i)
\end{equation},
where $a_i$ denotes the trade amount and the sign depends on the buy ($+$) or sell ($-$). The logarithmic considers the marginal utility to smooth the impact of extreme values.

\textbf{Party Weight ($w_{\text{pty}, i}$)}: This weight aims to transcend the limitations of subjective self-reporting, potential biases inherent in traditional polling or social media analyses, and reticent to public disclosure on political issues~\cite{dang2020self,jerit2020political,cowan2018could}. This weight is delineated as $\{+1, -1\}$, where $+1$ signifies bettings favoring Democratic candidates, $-1$ represents Republican. The determination directly hinges on the political affiliation of the candidate mentioned in the betting target within the presidential election winner 2024 event (We ignore the independent candidate, Robert F. Kennedy Jr.). 

\textbf{Frequency Weight ($w_{\text{freq}, u}$)}: Recent studies have elucidated the correlation between participation frequency and polarized attitude~\cite{argyle2022does}. The weight is conceptualized as
\begin{equation}
w_{\text{freq}, u} = \log(1 + |T_u|)
\end{equation},
where $|T_u|$ denotes the trade numbers of user $u$. While increased exposure to information shapes users' political beliefs and trading behaviors~\cite{demszky2019analyzing}, it is crucial to recognize the distinction between prediction markets and social media platforms. Frequent betting in prediction markets entails expressing an opinion and committing economic resources based on perceived probabilities, incorporating users' risk preferences. Hence, we introduced this weight and examined risk preference features in correlation analyses. 

In summary, each weight in the PBLS has its theoretical basis and practical considerations to balance theoretical soundness and practical operability. We apply the PBLS to a subdataset comprising 823 user addresses participating in the ``Presidential Election Winner 2024'' event\footnote{https://polymarket.com/event/presidential-election-winner-2024} on Polymarket. This enables a direct assessment of the political leanings manifested through betting behaviors.

\subsection{Validation}\label{method:pbls_val}
\textbf{Internal validation.} To validate the robustness, we first conducted an internal validation by comparing the PBLS derived from the presidential election with the PBLS calculated from another smaller concurrent event, which is predicting the 2024 popular vote\footnote{https://polymarket.com/event/presidential-election-popular-vote-winner-2024}. The internal validation aimed to assess the consistency of the PBLS across different markets within the same platform. For 106 user addresses who participated in both, we calculated their PBLS based on their bettings in each one separately. We then conducted a Kolmogorov-Smirnov test to assess whether the two PBLS distributions were significant difference. After that, we calculated their Pearson and Spearman correlation coefficients to measure the linear and rank-order relationship. 

\textbf{External validation}. We also validated externally by comparing PBLS with 173 times national poll data (326,695 samples) aggregated by ABC News regarding the 2024 U.S. presidential election, from the beginning of 2024 to February 27, 2024\footnote{https://projects.fivethirtyeight.com/polls/president-general/2024/national}. To align the PBLS with the binary polls, we classified users as Democratic or Republican supporters based on their PBLS and calculated the corresponding ratio. We then compared the support ratios with the weighted average ratio from the polls, where the weights were the sample size of each poll.    
Specifically, we employed a weighted bootstrap procedure to assess the statistical significance of this comparison. We resampled the polls with replacement, using the sample sizes as weights, and calculated the weighted average of the Democratic-to-Republican ratio for each bootstrap sample. This process was repeated 1,000 times to create a bootstrap distribution of the poll-based ratio. Then, we checked whether the PBLS-based Democratic-to-Republican ratio fell within the 95\% confidence interval of the real poll results distribution. The external validation aims to provide evidence for the construct validity of our PBLS. 

\subsection{Feature Engineering and Correlation}\label{method:fe_corr}
We conducted feature engineering to uncover factors influencing political leanings and enable PBLS calculation for users not participating in political markets. This process reveals potential determinants and enables widely measuring political leanings by leveraging the blockchain's ability.

We constructed three-level features: \textbf{user address}, \textbf{event}, and \textbf{market}, which describe user behavior patterns across different granularities. The features include eight categories: \textbf{Basic features} provide an overview of a user's engagement and performance. \textbf{Participation features} capture the distribution of user behaviors across different events and markets. \textbf{Trading behavior features} encompass users' buying and selling patterns. \textbf{Success rate and profitability features} assess predictive accuracy and financial performance. \textbf{Side preference features} capture beliefs and biases toward specific outcomes or propositions. \textbf{Time-related features} examine temporal patterns of user behaviors and their consistency around critical political issues. \textbf{Risk preference features} provide insights into risk diversification strategies and attention to specific segments. \textbf{Combination features} capture more subtle behavioral aspects. We obtained 825 features for subsequent analyses, which considered previous studies and prediction market mechanisms. The description and formula for each feature are detailed in Appendix~\ref{app:feature}.  

Then, we conducted a thorough correlation analysis to assess the relevance of these features to users' political leanings. Due to the diverse participation patterns of individual user addresses, some features were missing for users who did not engage in certain events or markets. To handle these missing values, we employed pairwise deletion, maximizing the use of available data. We calculated the Spearman correlation coefficients between the features and PBLS, as it is more robust to outliers and non-normal distributions. The complete correlation results are presented in Appendix~\ref{app:corr}. Among the 825 features, 533 exhibited significance with PBLS ($p$ $<$ 0.05), suggesting that these features contain valuable information for predicting users' political leanings.

\subsection{Predicting PBLS for All User Addresses}\label{method:ml_pred}
We employed machine learning to predict the PBLS of users who did not participate in the presidential election markets based on features from their behaviors in other markets.

We split the dataset into users with and without computed PBLS. The former was divided into training and testing sets (80\%-20\%). To select the most informative features, we applied Recursive Feature Elimination (RFE), which recursively removes features and builds models. To mitigate overfitting and ensure stability, we employed several strategies. First, we applied data augmentation with 0.01 scale random noise to standardized features, introducing slight perturbations to increase training data diversity and improve model generalization. Second, we compared the performance of different models, including linear regression, decision trees, and random forests, and found XGBoost~\cite{chen2016xgboost} performed best in predicting PBLS. Furthermore, we optimized hyperparameters using RandomizedSearchCV, including the number of estimators, learning rate, maximum depth, subsampling ratio, and regularization terms. The best model, selected based on 5-fold cross-validation and negative mean squared error, achieved an MSE of 0.415 and an $R^2$ of 0.644 on the testing set. Finally, we applied this optimized model to predict PBLS for all user addresses without it, effectively extending our measuring of political leanings to the entire dataset.

\subsection{Method for Case Study}\label{method:case_study}
To further investigate the relationship between political leanings and behaviors, we looked at the ``Which party will control the U.S. Senate after the 2022 election?''\footnote{https://polymarket.com/event/which-party-will-control-the-us-senate-after-the-2022-election?}, which ran from January 13, 2022, to November 14, 2022, focused on the Senate elections that began on November 8, 2022. Ultimately, the Democrats secured more than half seats (51:50). 

Our choice reasons include: 1. It finished with a complete cycle, available to track the evolution of political leanings and prices, providing a comprehension of their interplay on betting behaviors. 2. The targets, ``Democratic'' and ``Republican,'' are unambiguous and directly correspond to political leanings. 3. The 2022 U.S. midterm is significant, and the Senate election is the largest relevant event on Polymarket. 4. While Senate elections exhibit relative stability historically~\cite{lublin1994quality}, the recent outcomes diverged from mainstream poll predictions and academic studies~\cite{sabato2019blue,hummel2014fundamental}. In other words, external known information would less influence the betting process, allowing us to focus on the interplay effect. Notably, while this case study focuses on a specific U.S. political event, the design logic and methodology are generalizable.

\subsubsection{Visualizing PBLS and Profit/Loss} \label{method:case_study_pl}
First, we hypothesized that political leanings should be associated with betting behaviors in political-oriented prediction markets. Specifically, in this event where the Democratic Party ultimately won, we expected individuals with higher PBLS (i.e., more Democratic-leaning) would exhibit higher profits. To test this hypothesis, we visualized the distribution of users' PBLS alongside their realized profit and loss (P\&L) in the event.

To calculate each user address' P\&L, we first determined the cash flow (F) for each transaction $i$ based on the betting side (buy or sell) and the corresponding USDC size (uS):
\begin{equation}
\text{F}_i = \begin{cases}
-\text{uS}_i, & \text{if } \text{side}_i = \text{buy} \\
\text{uS}_i, & \text{if } \text{side}_i = \text{sell}
\end{cases}
\end{equation}

Next, we aggregated by address and outcome (Democratic or Republican, abbreviated as Dem and Rep, respectively) to calculate each user-outcome combination's total size and cash flow. Since the Dem won, we determined the final cash flow (FF) for each user-outcome combination as follows:
\begin{equation}
\text{FF}_{u,o} = \begin{cases}
\sum{i \in T_{u,o}} \text{size}_i, & \text{if } o = \text{Dem} \\
0, & \text{if } o = \text{Rep}
\end{cases}
\end{equation}
\noindent
, where $T_{u,o}$ is the set of betting for user address $u$ and outcome $o$. The total P\&L for each user was then calculated as the sum of their cash flow and final cash flow:
\begin{equation}
\text{P\&L}_u = \sum{o \in {(\text{Dem}, \text{Rep})}} (\sum_{i \in T_{u,o}} \text{F}_i + \text{FF}_{u,o})
\end{equation}

To assess the statistical significance of the relationship between PBLS and P\&L, we performed a regression analysis and constructed a 95\% confidence interval for the coefficient:
\begin{equation}
\text{P\&L}_u = \beta_0 + \beta_1 \text{PBLS}_u + \epsilon_u
\end{equation}
\noindent
, where $\beta_0$ and $\beta_1$ are the intercept and slope coefficients, and $\epsilon_u$ is the error term. A statistically significant positive $\beta_1$ coefficient would indicate that users with higher PBLS tend to have higher P\&L. The 95\% confidence interval for $\beta_1$ provides a range of plausible values for the true effect size.

\subsubsection{Panel Data Analysis of Price, PBLS, and Holding Size}\label{method:case_study_panel}
In prediction markets like Polymarket, betting behaviors are likely driven by two primary factors: desire to maximize profits and belief in the outcome~\cite{wolfers2004prediction,manski2006interpreting}. Prices, which reflect the collective expectations, serve as a proxy for the profitable motive, as traders seek to buy low and sell high~\cite{berg2008prediction}. On the other hand, the PBLS captures political beliefs and preferences, which may influence willingness to bet on certain outcomes regardless of the potential profits~\cite{forsythe1999wishes,cowgill2015corporate}. Therefore, our second interest in the case study focused on disentangling the effects of these two factors, prices and political leanings, on holding sizes for each outcome over time. We sought to determine whether these factors exerted independent or interactive effects on users' betting behavior and whether these effects varied temporally.

We constructed a panel dataset containing user addresses' holding sizes, the prices for each outcome, and the PBLS at multiple time points. We selected daily interval points before and after the election night (November 7, 2022) until the market's end (November 14, 2022). At each point, we determined the prices for the Dem and Rep outcomes based on the closest betting. For each address, we then calculated the cumulative holding size (hS) for each outcome by summing the sizes of transactions up to that time point:
\begin{equation}
\text{hS}_{u,o,t} = \sum{i \in \pm T_{u,o,t}} \text{size}_i
\end{equation}
\noindent
, where $T_{u,o,t}$ denotes the set of transactions for user $u$, outcome $o$, and time points up to and including $t$, and the sign depends on the buy ($+$) or sell ($-$). 

Then, we fit a random-effects model to estimate the effects of PBLS ($\alpha$), prices ($\beta$), and their interaction ($\gamma$) on holding size. A significant interaction term would indicate that the effect of prices on holding sizes varies depending on users' political leanings, and vice versa. The model is:
\begin{align}
\text{hS}{u,t} = (\theta_0 + \delta_u) + \theta_1 \alpha_u + \theta_2 \beta_t + \theta_3 \gamma{u,t} + \epsilon_{u,t}
\end{align}
\noindent
, where $\theta_0$, $\theta_1$, $\theta_2$, and $\theta_3$ are the fixed-effect coefficients, $\delta_u$ is the user-specific random intercept, and $\epsilon_{u,t}$ is the error term.

\begin{figure*}[!h]
\centering 
\begin{subfigure}{.331\textwidth}
  \centering
  \includegraphics[width=\linewidth]{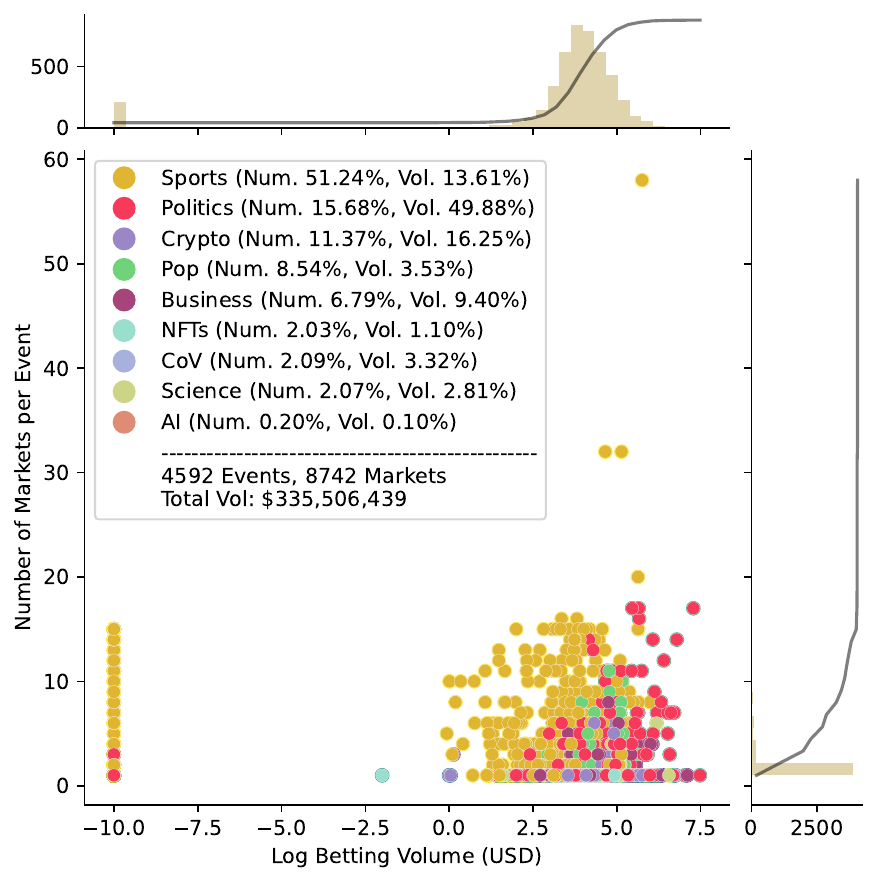}
  \caption{}
  \label{fig:result1-2}
\end{subfigure}
\begin{subfigure}{.331\textwidth}
  \centering
  \includegraphics[width=\linewidth]{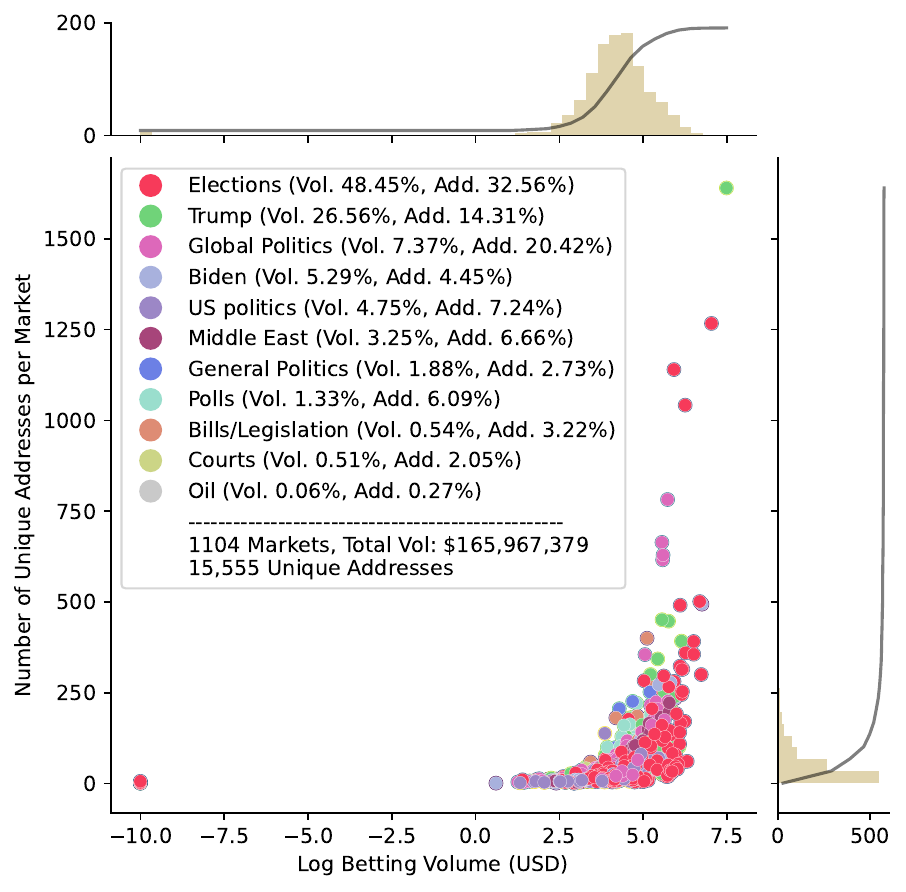}
    \caption{}
  \label{fig:result1-3}
\end{subfigure}%
\begin{subfigure}{.331\textwidth}
  \centering
  \includegraphics[width=\linewidth]{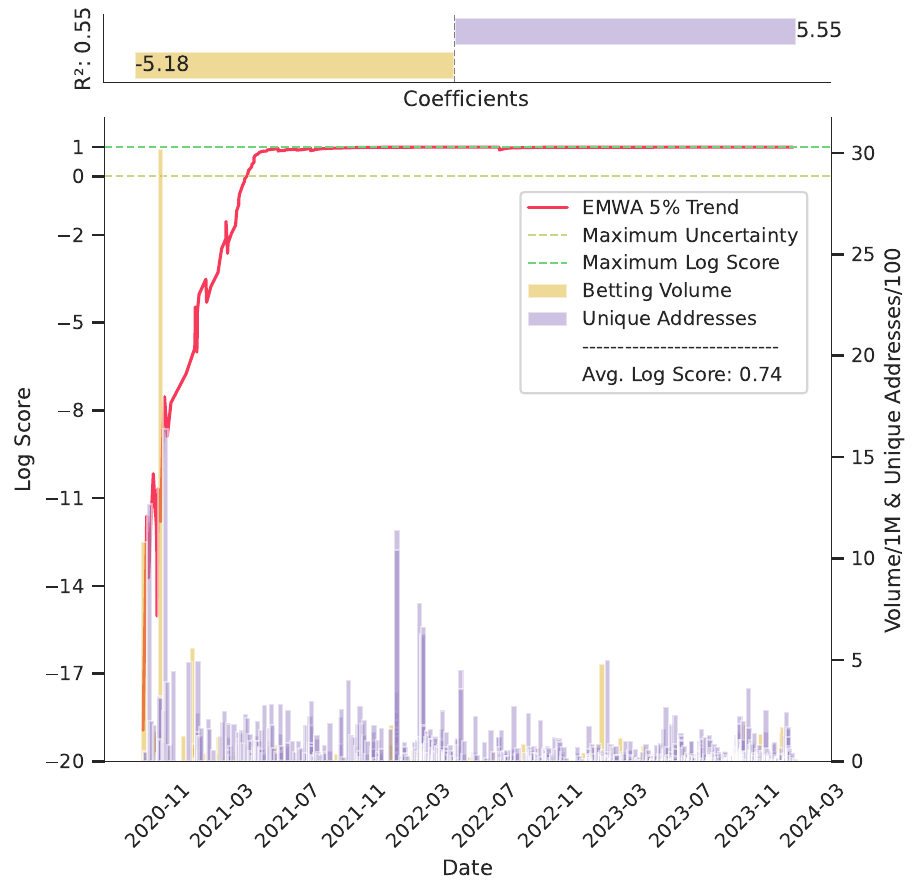}
    \caption{}
  \label{fig:result1-4}
\end{subfigure}
\caption{\textbf{(a)} Distribution of events. Politics-related events rank second in numbers but first in volume. The skewness of count and volume existed across events. \textbf{(b)} Distribution of politics-related markets. Elections markets attract the highest volume and unique user addresses. \textbf{(c)} The accuracy of political prediction markets improved and stabilized after an initial volatility, positively impacted by participation and negative by volume.}
\label{fig:result1}
\end{figure*}

\section{Results And Analysis}\label{result}
\subsection{Overview: Political Betting Dominance}\label{result:overview}
Our dataset covers 4,592 events and 8,742 markets on Polymarket from October 2, 2020, to February 27, 2024, with a total volume of \$335,506,439 (Section~\ref{method:data}). Figure~\ref{fig:result1-2} reveals that among the nine main event categories, ``Politics'' ranks second in the number (15.68\%) but first in betting volume (49.88\%), indicating significant user involvement. Moreover, most events have fewer than five markets, aligning with research on optimal crowd prediction design~\cite{brown2019wisdom}. Political events do not exhibit extreme numbers of markets, suggesting a balanced and potentially beneficial market structure.

Figure~\ref{fig:result1-3} presents a granular view of the political event category, which can be further classified into eleven market categories (\$165,967,379 total betting volume, 15,555 unique addresses). ``Elections'' markets are the most significant in betting volume (48.45\%) and participation (32.56\%), providing the foundation for our subsequent focus on specific election markets to investigate political leanings. While a single market related to the ``Trump'' category attracts the highest number of participants and volume, we opted not to center around it, as predictions concerning individual political figures may not necessarily reflect accurate attitudes~\cite{jottier2012understanding}. 

In summary, the salience of political betting on Polymarket supports our study of political leanings through this platform. The concentration on political events, especially election markets, enables a targeted analysis of how political leanings shape Web3 prediction market behaviors.

\subsection{Predictive Accuracy of Political Markets}\label{result:accuracy}
To evaluate the predictive accuracy of political markets on Polymarket, we employed methods in Section~\ref{method:accuracy}. As depicted in Figure~\ref{fig:result1-4}, the average DLS for political markets on Polymarket is 0.74, a little surpassing some famous counterparts, like Metaculus (0.645)~\cite{mann2016power} a collective prediction platform without blockchain ensured transparent access and aggregate. The EWMA line reveals that predictive accuracy was poor and volatile during Polymarket's early stages (early 2021) but improved and stabilized following this initial, aligning with the concept of prediction maturation, whereby the increased participant and information aggregation contribute to more efficient and accurate outcomes~\cite{berg2008prediction, strijbis2019explaining}.

To further investigate the factors influencing predictive accuracy, we analyzed the relationship between accuracy and two key market indicators at this stage: betting volume and unique address participants. The regression yielded an $R^2$ score of 0.5, indicating that these two variables collectively explain 50\% of the variance in predictive accuracy. The coefficients for betting volume and participants exhibit opposite signs (-5.18 and 5.55, respectively), suggesting that increased trading volume may introduce noise or speculation, while a larger participant improved accuracy, likely by incorporating a more diversity of information and perspectives~\cite{lang2016crowdsourcing}.

In summary, Polymarket's political markets exhibit high accuracy that improves over time, which more unique participants positively influence, while higher trading volume may introduce noise. These findings reveal the potential and complexity of blockchain prediction markets.

\begin{figure*}[htbp]
\centering 
\begin{subfigure}{.42\textwidth}
  \centering
  \includegraphics[width=.95\linewidth]{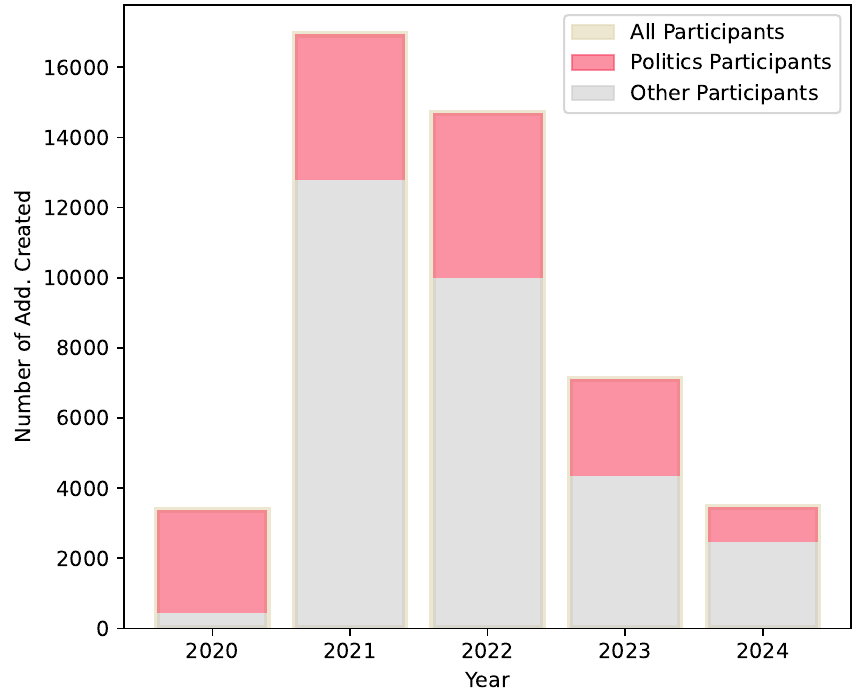}
    \caption{}
  \label{fig:result2-1}
\end{subfigure}%
\begin{subfigure}{.42\textwidth}
  \centering
  \includegraphics[width=.95\linewidth]{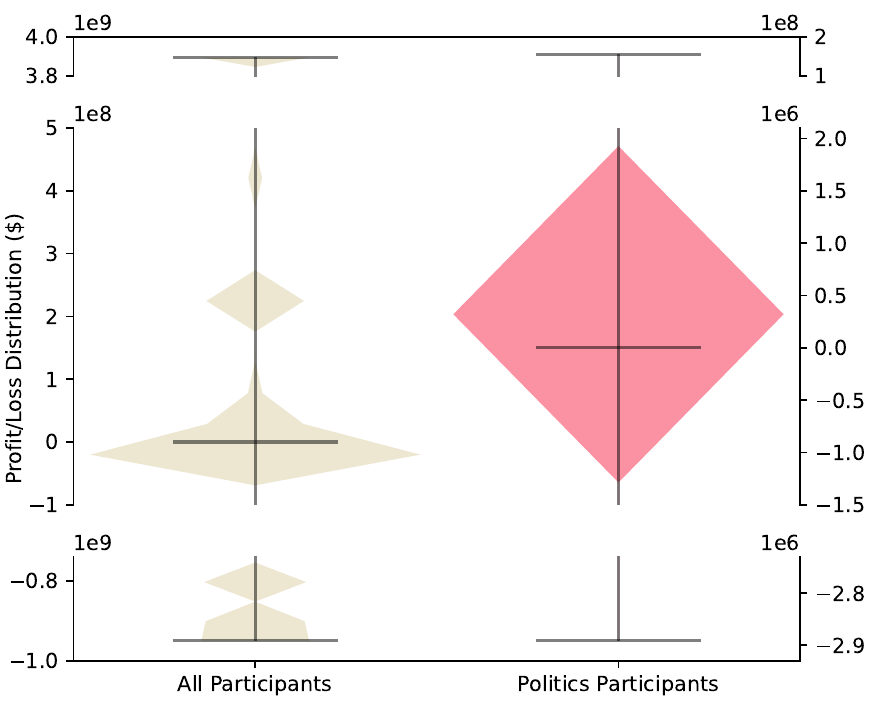}
    \caption{}
  \label{fig:result2-2}
\end{subfigure}

\begin{subfigure}{.42\textwidth}
  \centering
  \includegraphics[width=.95\linewidth]{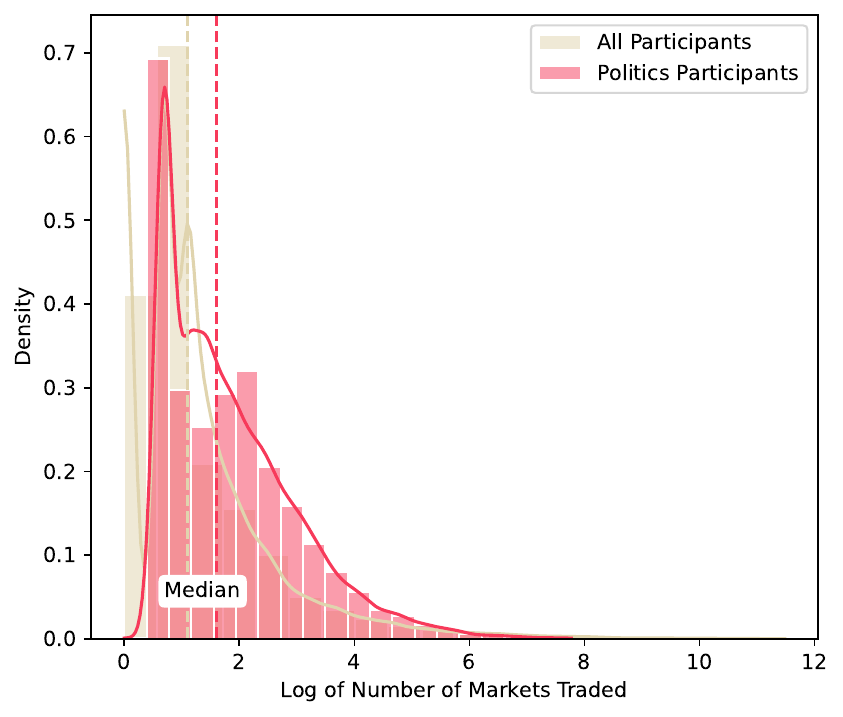}
    \caption{}
  \label{fig:result2-3}
\end{subfigure}%
\begin{subfigure}{.42\textwidth}
  \centering
  \includegraphics[width=.95\linewidth]{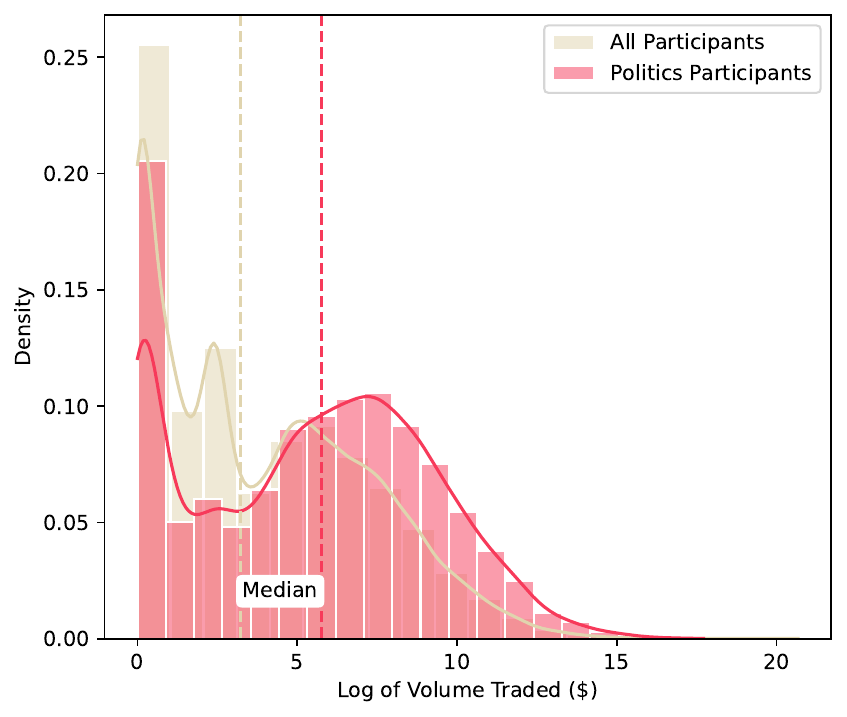}
    \caption{}
  \label{fig:result2-4}
\end{subfigure}
\caption{\textbf{(a)} Yearly user address creation on Polymarket: general vs. political market participants. \textbf{(b)} Realized Profit/Loss distribution: general vs. political betting participants. \textbf{(c)} Number of markets traded: general vs. political betting participants. \textbf{(d)} Volume traded: general vs. political betting participants.}
\label{fig:result2}
\end{figure*}

\subsection{Political V.S General Participants}\label{result:users}
Figure~\ref{fig:result2-1} shows distinctions in user address creation between the general Polymarket population and political betting participants. While new users peaked in 2021 for the general, likely driven by the bullish broader Web3 ecosystem, subsequently saw a decline, potentially due to the bear trend. However, the yearly creation of political market participants remained relatively stable, suggesting that political betting attracts a dedicated user base less susceptible to macro market fluctuations. This aligns with findings that political prediction often attracts intrinsically motivated participants who desire to express beliefs or gain reputational rewards rather than solely seeking financial returns~\cite{berg2008prediction,berg2019bet}.

Figure~\ref{fig:result2-2} reveals notable differences in realized profit and loss. The general population exhibits a distribution with multiple dense regions, a mean close to 0, and a negative median. The wide range between Q1 and Q3 suggests high variability in outcomes. In contrast, political betting participants display a more concentrated, spindle-shaped distribution, with a mean near 0 and a positive median, indicating a slightly higher likelihood of profitability and more consistency. Both 0 mean values could be partially explained by the zero-sum nature of betting markets, where the losses of others balance the profits of successful bettors. The higher median profitability among political bettors might indicate skill or information advantage compared to the general population~\cite{brown2012evidence}.

\begin{figure*}[t]
\centering 
\begin{subfigure}{.42\textwidth}
  \centering
  \includegraphics[width=.95\linewidth]{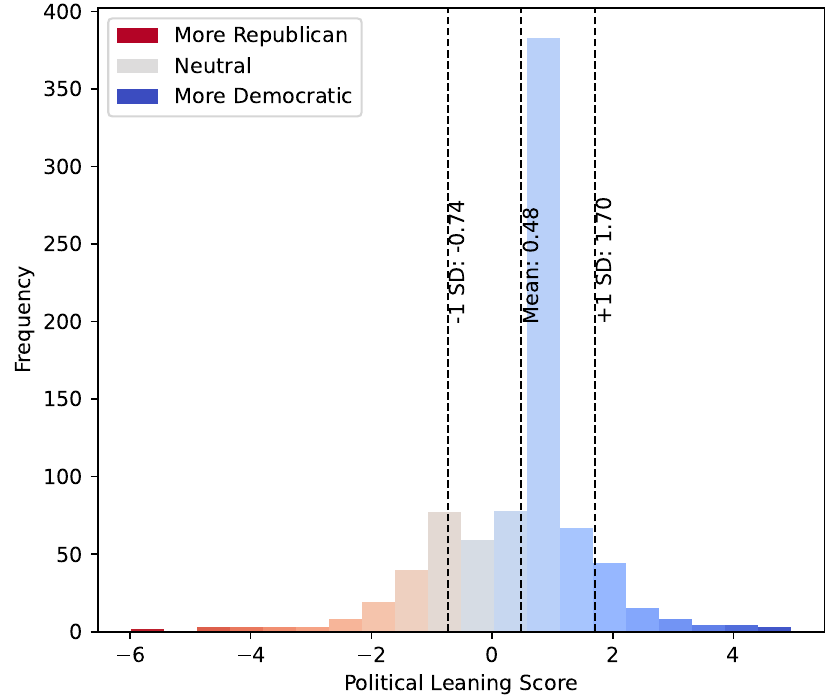}
    \caption{}
  \label{fig:result3-1}
\end{subfigure}%
\begin{subfigure}{.42\textwidth}
  \centering
  \includegraphics[width=.94\linewidth]{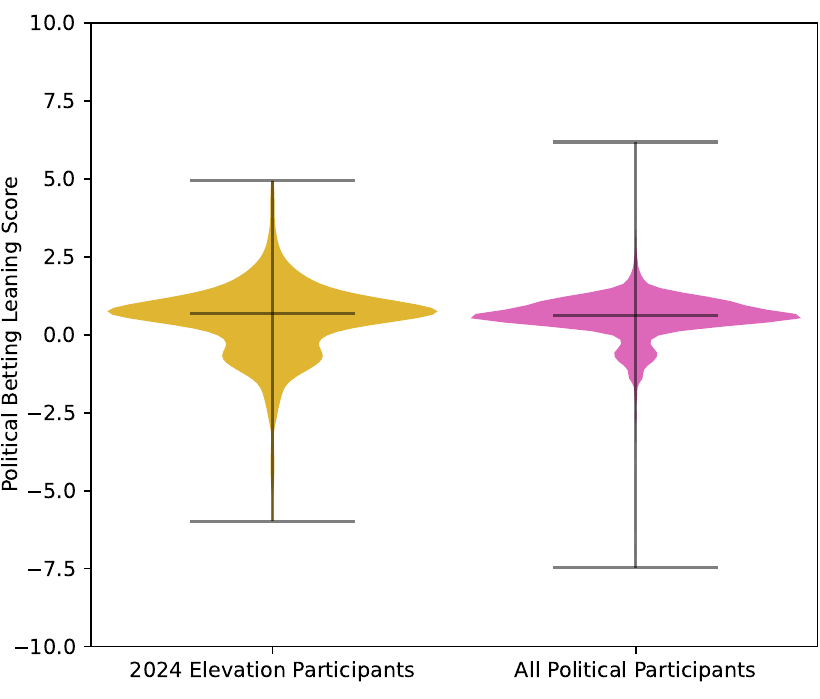}
    \caption{}
  \label{fig:result3-2}
\end{subfigure}

\begin{subfigure}{0.9\textwidth}
  \centering
  \includegraphics[width=\linewidth]{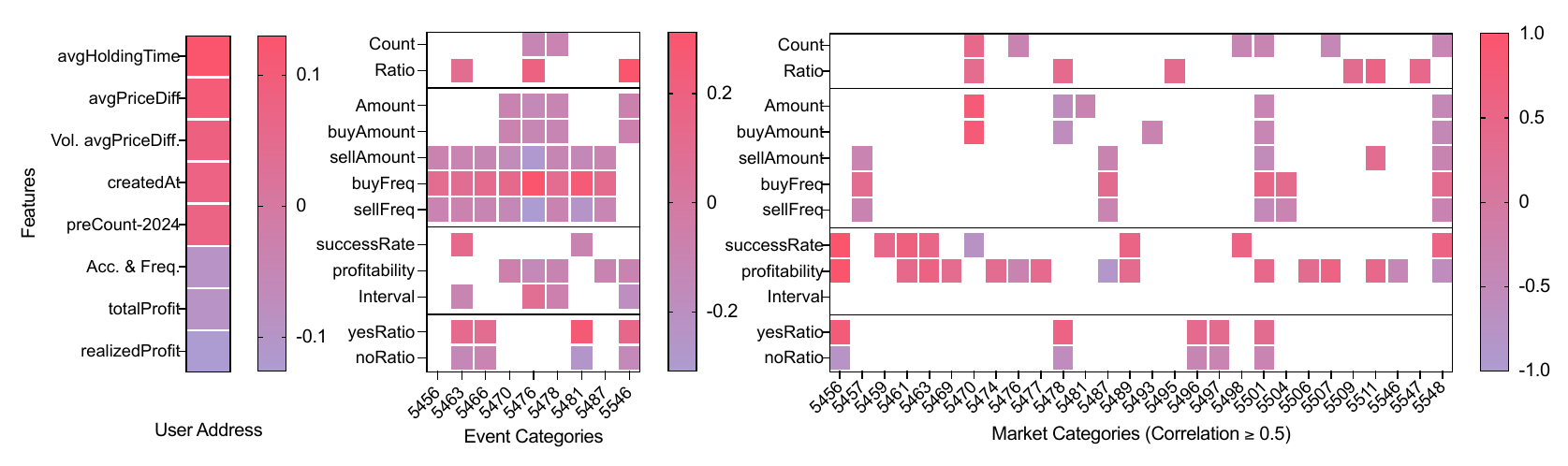}
    \caption{}
  \label{fig:result3-3}
\end{subfigure}%
\caption{\textbf{(a)} Distribution of Political Betting Leaning Scores (PBLS) for user addresses participated in the 2024 U.S. presidential election event on Polymarket, showing a majority with moderate partisan preferences. \textbf{(b)} Comparison of actual and predicted PBLS distributions reveals the effectiveness of the machine learning approach in capturing political leanings. \textbf{(c)} Correlation coefficients between features and PBLS increase from user to event to market levels, highlighting the importance of granular behavioral analysis in decoding political preferences.}
\label{fig:result3}
\end{figure*}

Figure~\ref{fig:result2-3} provides insights into engagement distinctions. A significant portion of the general population did not participate in betting after registration, while political participants were more consistent with a higher median and a shorter tail. This engaged pattern might reflect the specialized interests of political bettors, who may selectively participate in markets aligned with their knowledge or beliefs. Figure~\ref{fig:result2-4} shows that political participants tended to trade higher volumes but had fewer extremely high-volume traders than the general, suggesting more measured betting amounts.

In summary, the participants in political prediction markets exhibit distinct behavioral patterns, tending to be more consistently active, slightly more likely to be profitable, and engaging more selectively in markets, underscoring the feasibility of studying this sub-group to understand the interplay between political leanings and betting behaviors.

\subsection{Computing and Validating PBLS}\label{result:PBLS}
Our computed the Political Betting Leaning Score (PBLS) using the methodology detailed in Section~\ref{method:pbls}. This score quantifies the political leanings based on user betting behaviors in these markets, with positive values indicating a Democratic-leaning and negative indicating a Republican-leaning. The distribution of PBLS (Figure~\ref{fig:result3-1}) has a mean of 0.48 and a standard deviation of 1.22, exhibiting a bell-shaped curve with many users concentrated around the neutral (PBLS = 0), who could be considered ``median voters''~\cite{caplin1991aggregation}. This aligns with previous studies that most individuals are not strongly partisan and tend to cluster around the ideological center~\cite{gerber2012personality,huckfeldt1999accessibility}. Notably, the presence of these neutrals may also be attributed to their profit orientation, which we will discuss later. Interestingly, while more user addresses have a slightly Democratic-leaning PBLS, some Republican-leaning addresses display more vigorous intensity. This asymmetry is consistent with conservatives more prone to political polarization and ideological segregation online~\cite{boutyline2017social}.

To validate the robustness of PBLS, we conducted internal and external validations (Section~\ref{method:pbls_val}). For internal validation, the Kolmogorov-Smirnov test indicated no significant difference ($p$ = 0.059), suggesting that the PBLS was consistent across related events. The correlation coefficients were high and statistically significant (Pearson's $r$ = 0.627, $p$ $<$ 0.001; Spearman's $\rho$ = 0.534, $p$ $<$ 0.001). These findings demonstrate that the PBLS is a robust and consistent measure of users' political leaning, even when calculated from different markets within the same prediction platform. For external validation, the PBLS-based ratio (0.760) fell within the 95\% confidence interval of the poll-based ratio [0.703, 1.729], indicating that the PBLS is consistent with established measures of public opinion. This provides evidence for the construct validity~\cite{adcock2001measurement} of the PBLS as a measure method.

In summary, we introduced the PBLS to quantify political leanings based on betting behaviors and checked robustness and validity, paving the way for further analyses.

\subsection{Decoding PBLS Across Betting}\label{result:decodePBLS}
Based on our feature engineering and correlation analysis (Section~\ref{method:fe_corr}), we identified 533 significant features correlating with PBLS out of 825, spanning user address, event, and market levels. Figure~\ref{fig:result3-3} shows correlation coefficients increasing from user address to market levels, suggesting comprehensive aggregation of behaviors within various events and markets enabled by blockchain might uncover more robust signals of political orientations.

At the user level, many expected influential features show weak or few correlations, such as realized profit (-0.126**) (Table~\ref{tab:app-corr-user}), aligning with findings that individual features may not sufficiently capture complex behavioral patterns like political preferences~\cite{caprara2007personalization, verhulst2012correlation}. Moreover, features related to important time points, such as trades made before the 2022 midterm (preCount-2022, 0.053) and before the 2024 presidential election event (preCount-2024, 0.073*), demonstrate few or insignificant correlations, showing behavioral consistency while challenging the assumption that financial behaviors fluctuating around key political events~\cite{franzese2002electoral}.

Event-level analysis (Table~\ref{tab:app-corr-event}) reveals more pronounced correlations across various domains. The Politics event (ID 5481) correlates with buyFreq (0.251**), sellFreq (-0.251**), and yesRatio (0.261**), suggesting politically polarized users are more likely to buy into political markets and bet on the ``Yes'' side. The Coronavirus event (ID 5463) weakly correlates with successRate (0.127**), yesRatio (0.125**), and noRatio (-0.125**), indicating Democratic-leaning users may have been more accurate in predicting pandemic-related outcomes. For Web3-related events, the NFTs (ID 5476) and crypto (ID 5466) events show politically polarized users, despite being less likely to participate, tend to exhibit more active buying and selling behaviors when they engage.

The market level (Table~\ref{tab:app-corr-market}) uncovers starker correlations. The Elections market (ID 5482) has buyFreq (0.278**) and sellFreq (-0.278**), suggesting politically polarized users actively participate to express their leanings. Significant correlations are also found in various areas. The Esports market (ID 5493) exhibits correlations with Amount (-0.293**) and buyFreq (-0.306**), indicating distinct patterns among politically polarized users. The Coronavirus market (ID 5463) shows moderate correlations with successRate (0.464**) and strong with profitability (0.6**), consistent with the Event level. The Friend Tech market (ID 5548), a Web3-related one, has a moderate negative correlation with Amount (-0.438**) and a strong positive with successRate (0.569**), suggesting Democratic-leaning users trade in smaller amounts but more accurately, and vice versa. Interestingly, some indirect markets, like Cricket (ID 5501), shows moderate correlations with sellFreq (-0.466**), buyFreq (0.466**), and profitability (0.45**), indicating distinctive strategies and profitability of politically leaned users in various domains.

In summary, due to the transparency of blockchain data, we obtained detailed user behaviors for in-depth and comprehensive analysis, which is difficult or costly in traditional prediction markets. Our feature engineering process reveals the complex patterns in how political leanings manifest in betting activities, supporting subsequent predictions.

\subsection{Predicting PBLS}
We predicted the PBLS using the approach detailed in Section \ref{method:ml_pred} to extend our analysis on the Polymarket user base. Figure~\ref{fig:result3-2} compares the actual PBLS, calculated for users who directly participated in the presidential election event, and the predicted PBLS of all political market participants.

\begin{figure*}[htbp]
\centering 
\begin{subfigure}{.45\textwidth}
  \centering
  \includegraphics[width=.95\linewidth]{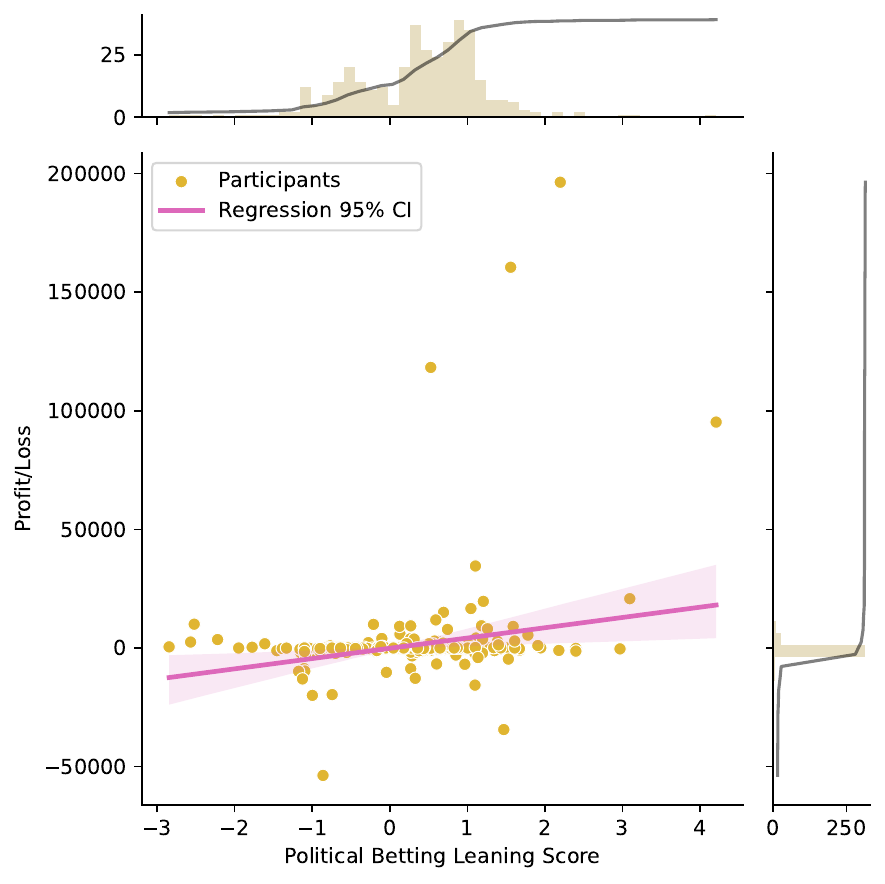}
    \caption{}
  \label{fig:case_1}
\end{subfigure}%
\begin{subfigure}{.45\textwidth}
  \centering
  \includegraphics[width=.95\linewidth]{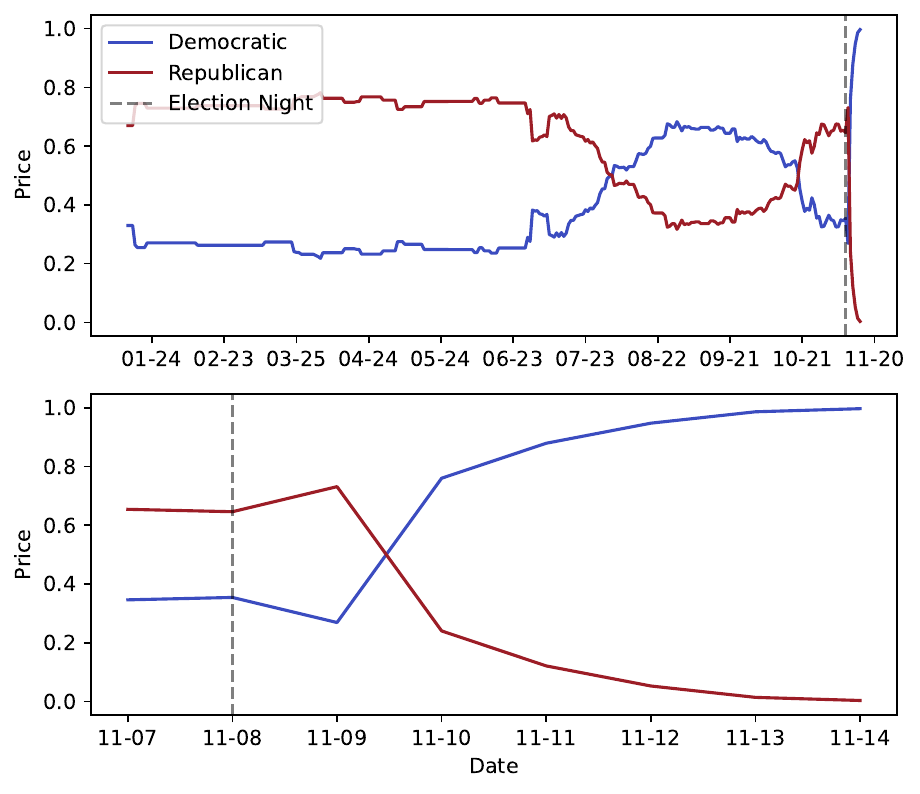}
    \caption{}
  \label{fig:case_2}
\end{subfigure}
    \caption{\textbf{(a)} The distribution of users’ PBLS alongside their realized profit and loss (P/L) in the event. \textbf{(b)} Price dynamics for the Democratic and Republican outcomes in the 2022 U.S. Senate election on Polymarket. The upper panel shows the price trends during the total event, indicating a long-term advantage for the Republican party before the election night (November 8, 2022). The lower panel presents the situation after the election night, revealing a shift towards the Democratic party. The vertical dashed line in both panels marks the election night.}
    \label{fig:case}
\end{figure*}

Firstly, both distributions resemble a spindle shape, with most users concentrated around the neutral center (PBLS close to 0) and tapering off towards the more extreme ends of the political spectrum. This suggests that while Polymarket attracts users across the political divide, most participants tend to exhibit moderate leanings. However, we observe a more pronounced concentration of users in the neutral for the predicted PBLS distribution than the actual PBLS distribution. This indicates that our machine learning approach may be conservative in assigning extreme PBLS values, possibly due to the inherent complexity of capturing the full spectrum of political preferences from betting behaviors. Comparing the tails of the distributions, we find that the predicted PBLS distribution has a higher proportion of users with extreme leanings relative to the actual PBLS distribution. This implies that by considering the full spectrum of users' betting behaviors across various markets, our machine learning approach can identify more users with strong partisan preferences that may not be evident when focusing solely on their participation in the presidential election.

In summary, prediction PBLS extends our analysis to all political market participants on Polymarket. These quantitative results support our assumption that political leanings can be decoded by harnessing the rich, granular blockchain behavioral data on decentralized prediction platforms.

\subsection{Case Study: 2022 U.S. Senate Election}\label{result:case_study}
\textbf{Visualizing PBLS and Profit/Loss}. In this politically oriented prediction, we hypothesized that political leanings would be associated with betting behaviors. As the Democrats ultimately won, we expected individuals with higher PBLS (i.e., more Democratic-leaning) to exhibit higher profits. To test this hypothesis (See Section~\ref{method:case_study_pl}), we visualized the distribution of PBLS alongside their realized profit and loss (P\&L) in the event (Fig.~\ref{fig:case_1}).

The regression in the scatter plot revealed a significant positive relationship between PBLS and P\&L, suggesting that Democratic-leaning bettors might have better judgment. The marginal distributions showed that most PBLS scores were concentrated around 0, consistent with most participants having relatively moderate political leanings. The P\&L distribution revealed that while most users had modest profits, there were some significant winners, with a few outliers having substantial profits or losses. Despite the overall positive correlation between PBLS and P\&L, some bettors with high PBLS still suffered losses or had minimal profits, implying that factors beyond political leaning, such as market factors, also influenced betting decisions and outcomes.

\textbf{Panel Data Analysis of Price, PBLS, and Holding Size}. Participants' bettings are influenced mainly by their profit goals and outcome beliefs. Prices are a proxy for profit motives, with aims to buy low and sell high. Conversely, the PBLS reflects their political beliefs, possibly affecting betting willingness on specific outcomes, independent of profit potential. Our case study's second focus was to analyze the separate and interactive impacts of prices and political leanings on users' bet sizes over time.

As Figure~\ref{fig:case_2}, the advantage shifted throughout the event period. Before the election night, the Republican generally had the advantage. However, it flipped after the election night (November 8, 2022) since the outcome gradually became clear, with the Democrats taking the lead and eventually securing the majority in the Senate. Hence, we constructed a daily panel dataset with election night as the splitter, until the event's end. We then fit a random-effects model to estimate the effects of PBLS, prices, and their interaction on users' holding sizes for each outcome (see Section~\ref{method:case_study_panel}). 

\begin{table}[htbp]
\centering
\renewcommand\arraystretch{1.2}
\resizebox{0.95\linewidth}{!}{
\begin{tabular}{c|cc|cc|cc}
\toprule
\multirow{2}{*}{\begin{tabular}[c]{@{}c@{}}\textbf{Dep. Variable}\\ Holding Size\end{tabular}} & \multicolumn{2}{c}{\textbf{Total Market}} & \multicolumn{2}{c}{\textbf{Before Election Night}} & \multicolumn{2}{c}{\textbf{After Election Night}} \\ \cline{2-7}      & Dem            & Rep             & Dem                 & Rep                 & Dem                & Rep                 \\ \cline{1-1}
\textbf{Const}                           & 964.79         & 2394.5***       & 1100.6              & 1290.5              & 1569.8             & 3142.7***           \\
\textbf{PBLS}                            & 1542.6         & -659.56         & 3464.3*             & -1186.6**           & 4826.1**           & -266.38             \\
\textbf{Price}                           & 587.85*        & -953.42***      & -42.489             & 399.28**            & 278.78             & -1702.5***          \\
\textbf{PBLS $\times$ Price} & 3452.7***      & 279.45*         & 1360.5***           & 1116.3***           & 1777.1***          & -502.22             \\ 
\bottomrule
\multicolumn{5}{l}{\footnotesize{$^{*}p<0.05$, $^{**}p<0.01$, $^{***}p<0.001$}} \\
\end{tabular}
}
\caption{Summary of Panel Data Analyses}
\label{tab:case_2}
\end{table}

Table~\ref{tab:case_2} shows the key coefficients from the panel data analyses (Appendix~\ref{app:panel}). For the total period, although the PBLS is not significant, the interaction between PBLS and price is significant and positive for both Democratic and Republican holding sizes, suggesting that political leanings do not directly influence holding sizes but rather impact the effect of price changes on holdings. In other words, users with different political preferences respond differently to price fluctuations. Moreover, the interaction effect is much larger for the Democratic side (3452.7***) than the Republican (279.45*), indicating an asymmetric influence. The price coefficient is positive for Democratic holdings (587.85*) and negative for Republican (-953.42***), implying that price increases lead to more buying among Democratic-leaning users but more selling among Republican-leaning users. 

Comparing the results before and after the election night reveals further details. Before the night, the PBLS effect was significant for both sides. The price effect was 399.28** for the leading party's (Republican) holdings but insignificant for the trailing (Democratic). The interaction was similarly significant for both. These suggest that before the night, when the outcome was still uncertain, the trailing supporters were primarily driven by their political beliefs, while both their beliefs and prices influenced the leading party supporters. 

After the night, the advantage flipped. the PBLS (4826.1**) and interaction (1777.1***) effects were significant for the leading party's (Democratic) holdings, while the price was insignificant. These indicate that as the outcome gradually becomes clearer, the leading party supporters were primarily driven by their political beliefs and were not engaging in conspicuous trend-following purchases. However, the PBLS effect became insignificant, and the price was -1702.5** for Republicans, indicating a shift in Republican-leaning users' behavior after they lost the advantage, appeared to abandon their political convictions and focus on minimizing losses, a transition from mix- to profit-oriented. The insignificant interaction effect implies that profit-oriented behavior was prevalent among the trailing supporters at this period and did not depend on their political leaning degrees.

Panel data analyses highlight the complex interplay between political and profitable motives. While initial positions were largely determined by political leanings, subsequent decisions were influenced by a combination of price movements and partisan preferences, with asymmetric effects for different sides. The results also underscore the temporal dynamics of these effects, as the relative influence of political leanings and prices may shift with the advantage changes.

\section{Discussion and Conclusion}\label{discussion}
We introduce a novel method for measuring political leanings through betting behaviors on the Web3 prediction market Polymarket. By constructing the Political Betting Leaning Score (PBLS) and conducting extensive feature engineering and machine learning analyses, we demonstrate the feasibility and effectiveness of leveraging blockchain data to gain insights into the complex interplay between users' political and economic motives in decentralized online environments.

Our findings highlight the value of blockchain-based prediction markets as a unique data source for understanding online user behavior. Its transparent, immutable, and granular nature enables researchers to capture users' indirectly revealed preferences and study their behavior detailedly, which is often challenging to achieve on traditional online platforms. The PBLS, validated through internal consistency checks and external comparisons with polls, showcases the potential of using on-chain betting data to quantify users' political leanings objectively and comprehensively.

Moreover, our case study on the 2022 Senate election provides valuable insights into the dynamic between political beliefs and market trends. The significant positive relationship between PBLS and profit/loss suggests that political leanings can translate into tangible economic outcomes in prediction markets. However, the panel data analysis reveals a more nuanced picture, indicating that the relative influence of political preferences and price fluctuations on betting behavior varies depending on the market phase and the perceived advantage of each party. These findings underscore the importance of considering both political and economic motives when studying user behavior in prediction platforms and highlight the temporal dynamics of these effects.

This study has important implications for the design and optimization of online prediction markets. By understanding how political and profitable motives shape trading strategies, platform developers can create more informed incentive mechanisms and market rules to encourage diverse participation, improve information aggregation, and mitigate potential biases. Additionally, the PBLS and the feature engineering process presented in this work can serve as a foundation for developing more sophisticated user profiling and behavioral modeling techniques, which could be applied to a wide range of online platforms beyond prediction markets.

However, it is essential to acknowledge the limitations and the potential challenges in generalizing our findings. First, the Polymarket user base may not be representative of the general population, as users of blockchain-based platforms tend to be more technologically savvy, risk-tolerant, and politically engaged than the average~\cite{chen2022digital,weyl2022decentralized}. Future research could incorporate data from other prediction markets and online platforms and compare the findings across different populations and cultural contexts.

Second, while we have provided theoretical justifications for the weights and parameters used in constructing the PBLS, further validation and optimization may be necessary. Future studies could explore alternative weighting schemes, incorporate more diverse data sources, and employ more advanced machine learning techniques to improve the robustness and predictive power of the PBLS. Additionally, researchers could investigate the potential of combining blockchain data with other data types, such as volunteer surveys, to obtain a more comprehensive understanding of user behavior and preferences.

Third, our study primarily focuses on the U.S. political context, and the generalizability of our findings to other countries and political systems remains to be explored. Future research could adapt the PBLS and the feature engineering process to different cultural and institutional settings and examine the cross-cultural variability in the relationship between political leanings and betting behavior.

In conclusion, this study shows the value of blockchain data in uncovering the intricate relationship between political preferences and economic motives in online prediction markets. By introducing the PBLS and conducting a comprehensive analysis of user behavior on Polymarket, we contribute to the growing body of literature on measuring decentralized platforms' role in shaping online user behavior and decision-making. Our findings highlight the importance of interdisciplinary approaches and data-driven methods in uncovering increasingly complex online behaviors and provide valuable insights for researchers, platform designers, and policymakers, etc. As blockchain technology continues to evolve and massive adoption, we believe that studies like this will play a crucial role in harnessing its potential for social-technical research and fostering a more informed, responsible, and empowering digital future.

\bibliographystyle{ACM-Reference-Format}
\bibliography{acmart}

\appendix

\section{Performance of Fine-tuning}\label{app:fine}
Due to missing category IDs for some markets in the Polymarket API, we fine-tuned the Babbage-002 model\footnote{https://platform.openai.com/docs/models} using the names and corresponding IDs of 6,097 markets and subsequently employed the fine-tuned model to fill in the missing IDs. Table~\ref{tab:fine-tuned} shows its performance and you can find related codes in our online resources. Considering that ID filling is a multi-class classification and that market names have diverse expressions and ID classification might be subjective, the performance is acceptable. In future work, more powerful models could be the alternative choice.
\begin{table}[htbp]
\centering
\renewcommand\arraystretch{0.75}
\resizebox{0.45\linewidth}{!}{ 
\begin{tabular}{c|c}
\toprule
\textbf{Model}     & babbage-002 \\ \midrule
\textbf{Accuracy}  & 0.743       \\
\textbf{Precision} & 0.745       \\
\textbf{Recall}    & 0.743       \\
\textbf{F1}        & 0.725       \\ \bottomrule
\end{tabular}
}
\caption{The Performance of Fine-tuned Model}
\label{tab:fine-tuned}
\end{table}

\section{Feature Engineering}\label{app:feature}
\textbf{Basic features}: These include account creation time ($F_{\text{createdAt}}\\ = T_{user}$), total realized profit ($F_{\text{realizedProfit}} = P_{user}$), number of events or markets traded ($F_{\text{eventsTraded}} = E_{user}$, $F_{\text{marketsTraded}} = M_{user}$), total trading volume ($F_{\text{volumeTraded}} = V_{user}$), and total activity count ($F_{\text{activityNum}} = A_{user}$). These fundamental metrics provide an overview of a user's engagement and performance on the platform.

\textbf{Participation Features}: We captured the distribution of user activities across different categories (event and market). For each category $i$, we calculated the count ($F_{\text{categoryCount},i}$) and ratio ($F_{\text{categoryRatio},i}$) of user activities. These features reveal a user's participation and familiarity in various domains:
\begin{equation}
F_{\text{categoryCount},i} = \sum_{j=1}^{A_{user}} \mathbf{1}{K_j = i}
\end{equation}
\begin{equation}
F_{\text{categoryRatio},i} = \frac{F_{\text{categoryCount},i}}{A_{user}}
\end{equation}
, where $i$ represents different categories (event or market), $K_j$ is the category of activity $j$, and $\mathbf{1}$ is the indicator.

\textbf{Trading behavior features}: These features encompass the buying and selling patterns, including the total amount bought ($F_{\text{buyAmount},i}$) and sold ($F_{\text{sellAmount},i}$), as well as the buying ($F_{\text{buyFrequency},i}$) and selling ($F_{\text{sellFrequency},i}$) frequencies for each event and market category $i$. They are related to a user's trading strategies and decision-making processes:
\begin{equation}
F_{\text{buyAmount},i} = \sum_{j=1}^{A_{user}} \mathbf{1}{K_j = i} \cdot \mathbf{1}{S_j = Buy} \cdot U_j
\end{equation}
\begin{equation}
F_{\text{sellAmount},i} = \sum_{j=1}^{A_{user}} \mathbf{1}{K_j = i} \cdot \mathbf{1}{S_j = Sell} \cdot U_j
\end{equation}
\begin{equation}
F_{\text{buyFrequency},i} = \frac{\sum_{j=1}^{A_{user}} \mathbf{1}{K_j = i} \cdot \mathbf{1}{S_j = Buy}}{F_{\text{categoryCount},i}}
\end{equation}
\begin{equation}
F_{\text{sellFrequency},i} = \frac{\sum_{j=1}^{A_{user}} \mathbf{1}{K_j = i} \cdot \mathbf{1}{S_j = Sell}}{F_{\text{categoryCount},i}}
\end{equation}
, where $i$ represents different categories (event or market), $K_j$ is the category of activity $j$, $S_j$ is the side (buy or sell) of activity $j$, $U_j$ is the amount of activity $j$, and $\mathbf{1}$ is the indicator.

\textbf{Success rate and profitability features}: We assessed user addresses' predictive accuracy and financial performance by calculating the success rate ($F_{\text{successRate},i}$) and profitability ($F_{\text{profitability},i}$) for each event and market category $i$. These features provide insights into a user's ability to forecast outcomes and generate returns:
\begin{equation}
F_{\text{successRate},i} = \frac{\sum_{j=1}^{A_{user}} \mathbf{1}{K_j = i} \cdot \mathbf{1}{O_j = R_j}}{F_{\text{categoryCount},i}}
\end{equation}
\begin{equation}
F_{\text{profitability},i} = \sum_{j=1}^{A_{user}} \mathbf{1}{K_j = i} \cdot (H_j \cdot \mathbf{1}{R_j = Yes} - H_j \cdot \mathbf{1}_{R_j = No})
\end{equation}
, where $i$ represents different categories (event or market), $K_j$ is the category of activity $j$, $O_j$ is the predicted outcome of activity $j$, $R_j$ is the actual outcome of activity $j$, $H_j$ is the holding at the end of activity $j$, and $\mathbf{1}$ is the indicator.

\textbf{Side preference features}:
For each event and market category $i$, we calculated the ratio of the amount traded on the ``Yes'' side ($F_{\text{yesRatio},i}$) and the ``No'' side ($F_{\text{noRatio},i}$) to capture the preferences for different outcomes. These features provide insights into users addresses' beliefs and biases towards specific event outcomes or market propositions:
\begin{equation}
F_{\text{yesAmount},i} = \sum_{j=1}^{A_{user}} \mathbf{1}{K_j = i} \cdot \mathbf{1}{O_j = Yes} \cdot U_j
\end{equation}
\begin{equation}
F_{\text{noAmount},i} = \sum_{j=1}^{A_{user}} \mathbf{1}{K_j = i} \cdot \mathbf{1}{O_j = No} \cdot U_j
\end{equation}
\begin{equation}
F_{\text{yesRatio},i} = \frac{F_{\text{yesAmount},i}}{F_{\text{yesAmount},i} + F_{\text{noAmount},i}}
\end{equation}
\begin{equation}
F_{\text{noRatio},i} = \frac{F_{\text{noAmount},i}}{F_{\text{yesAmount},i} + F_{\text{noAmount},i}}
\end{equation}
, where $i$ represents different categories (event or market), $K_j$ is the category of activity $j$, $O_j$ is the outcome (Yes or No) of activity $j$, $U_j$ is the amount of activity $j$, and $\mathbf{1}$ is the indicator.

\textbf{Time-related features}: To capture temporal patterns, we computed the average holding time ($F_{\text{avgHoldingTime}}$) and average trading interval ($F_{\text{avgTradingInterval}}$) for each user address. We also set two date points: the start of the U.S. 2022 midterm elections (November 8, 2022) and the 2024 presidential election event on Polymarket (January 4, 2024). We then calculated the number of trades and average trading amounts before ($F_{\text{preCount}}$, $F_{\text{preAmount}}$) and after ($F_{\text{postCount}}$, $F_{\text{postAmount}}$) these points separately. The choice of these two observation points is grounded in their political significance and potential impact on user behavior. The U.S. midterm elections in 2022 served as a critical juncture during our dataset, with the outcomes influencing the balance of power in Congress. The start of the 2024 presidential election on Polymarket marks the beginning of a new election cycle. By comparing user behaviors before and after these dates, we examine the consistency of political leanings:
\begin{equation}
F_{\text{avgHoldingTime}} = \frac{\sum_{j=1}^{A_{user}} (T_{j,sell} - T_{j,buy}) \cdot U_j}{\sum_{j=1}^{A_{user}} U_j}
\end{equation}
\begin{equation}
F_{\text{avgTradingInterval}} = \frac{\sum_{j=2}^{A_{user}} (T_j - T_{j-1})}{A_{user} - 1}
\end{equation}
\begin{equation}
F_{\text{preCount}} = \sum_{j=1}^{A_{user}} \mathbf{1}{T_j < T{event}}
\end{equation}
\begin{equation}
F_{\text{postCount}} = \sum_{j=1}^{A_{user}} \mathbf{1}{T_j \geq T{event}}
\end{equation}
\begin{equation}
F_{\text{preAmount}} = \frac{\sum_{j=1}^{A_{user}} \mathbf{1}{T_j < T{event}} \cdot U_j}{F_{\text{preCount}}}
\end{equation}
\begin{equation}
F_{\text{postAmount}} = \frac{\sum_{j=1}^{A_{user}} \mathbf{1}{T_j \geq T{event}} \cdot U_j}{F_{\text{postCount}}}
\end{equation}

\begin{table*}[htbp]
\centering
\renewcommand\arraystretch{0.7}
\resizebox{0.6\linewidth}{!}{ 
\begin{tabular}{cccccc}
\toprule
\begin{tabular}[c]{@{}c@{}}\textbf{created}\\ \textbf{At}\end{tabular}          & \begin{tabular}[c]{@{}c@{}}\textbf{total}\\ \textbf{Profit}\end{tabular}         & \begin{tabular}[c]{@{}c@{}}\textbf{realized}\\ \textbf{Profit}\end{tabular} & \begin{tabular}[c]{@{}c@{}}\textbf{events}\\ \textbf{Traded}\end{tabular}      & \begin{tabular}[c]{@{}c@{}}\textbf{markets}\\ \textbf{Traded}\end{tabular}      & \begin{tabular}[c]{@{}c@{}}\textbf{volume}\\ \textbf{Traded}\end{tabular}            \\ 
0.074*                                                        & -0.097**                                                       & -0.126**                                                  & -0.009                                                       & 0.003                                                         & 0.01                                                               \\ \midrule
\begin{tabular}[c]{@{}c@{}}\textbf{activity}\\ \textbf{Num}\end{tabular}        & \begin{tabular}[c]{@{}c@{}}\textbf{avgHolding}\\ \textbf{Time}\end{tabular}      & \begin{tabular}[c]{@{}c@{}}\textbf{preCount}\\ \textbf{-2022}\end{tabular}  & \begin{tabular}[c]{@{}c@{}}\textbf{postCount}\\ \textbf{-2022}\end{tabular}    & \begin{tabular}[c]{@{}c@{}}\textbf{preAmount}\\ \textbf{-2022}\end{tabular}     & \begin{tabular}[c]{@{}c@{}}\textbf{postAmount}\\ \textbf{-2022}\end{tabular}         \\ 
0.01                                                          & 0.13**                                                         & 0.053                                                     & -0.002                                                       & 0.008                                                         & 0.007                                                              \\ \midrule
\begin{tabular}[c]{@{}c@{}}\textbf{preCount}\\ \textbf{-2024}\end{tabular}      & \begin{tabular}[c]{@{}c@{}}\textbf{postCount}\\ \textbf{-2024}\end{tabular}      & \begin{tabular}[c]{@{}c@{}}\textbf{preAmount}\\ \textbf{-2024}\end{tabular} & \begin{tabular}[c]{@{}c@{}}\textbf{postAmount}\\ \textbf{-2024}\end{tabular}   & \begin{tabular}[c]{@{}c@{}}\textbf{avgPrice}\\ \textbf{Diff}\end{tabular}       & \begin{tabular}[c]{@{}c@{}}\textbf{volWeighted}\\ \textbf{AvgPriceDiff}\end{tabular} \\ 
0.073*                                                        & -0.027                                                         & -0.032                                                    & 0.009                                                        & 0.102**                                                       & 0.084*                                                             \\ \midrule
\begin{tabular}[c]{@{}c@{}}\textbf{event}\\ \textbf{Concentration}\end{tabular} & \begin{tabular}[c]{@{}c@{}}\textbf{market}\\ \textbf{Concentration}\end{tabular} & \begin{tabular}[c]{@{}c@{}}\textbf{avgTrade}\\ \textbf{Amount}\end{tabular} & \begin{tabular}[c]{@{}c@{}}\textbf{avgAmount}\\ \textbf{PerEvent}\end{tabular} & \begin{tabular}[c]{@{}c@{}}\textbf{avgAmount}\\ \textbf{PerMarket}\end{tabular} & \begin{tabular}[c]{@{}c@{}}\textbf{accFreq}\\ \textbf{Product}\end{tabular}          \\ 
-0.027                                                        & -0.031                                                         & 0.016                                                     & 0.014                                                        & 0.011                                                         & -0.096**                                                          
\\ \bottomrule           
\end{tabular}
}
\caption{Correlation Analysis Results: User-level Features} 
\label{tab:app-corr-user}
\end{table*}

\begin{table*}[htbp]
\centering
\renewcommand\arraystretch{0.8}
\resizebox{0.9\linewidth}{!}{ 
\begin{tabular}{cc|ccccccccccc}
\toprule
\textbf{ID}   & \textbf{Name}     & \begin{tabular}[c]{@{}c@{}}\textbf{category}\\ \textbf{Count}\end{tabular} & \begin{tabular}[c]{@{}c@{}}\textbf{category}\\ \textbf{Ratio}\end{tabular} & \textbf{Amount}   & \begin{tabular}[c]{@{}c@{}}\textbf{buy}\\ \textbf{Amount}\end{tabular} & \begin{tabular}[c]{@{}c@{}}\textbf{sell}\\ \textbf{Amount}\end{tabular} & \begin{tabular}[c]{@{}c@{}}\textbf{buy}\\ \textbf{Freq}\end{tabular} & \begin{tabular}[c]{@{}c@{}}\textbf{sell}\\ \textbf{Freq}\end{tabular} & \begin{tabular}[c]{@{}c@{}}\textbf{success}\\ \textbf{Rate}\end{tabular} & \textbf{profitability} & \begin{tabular}[c]{@{}c@{}}\textbf{yes}\\ \textbf{Ratio}\end{tabular} & \begin{tabular}[c]{@{}c@{}}\textbf{no}\\ \textbf{Ratio}\end{tabular} \\ \midrule
5456 & Business & -0.003                                                   & -0.045                                                   & -0.004   & 0.006                                                & -0.084*                                               & 0.099**                                            & -0.099**                                            & -0.034                                                 & -0.049        & 0.049                                               & -0.049                                             \\
5463 & Coron    & -0.04                                                    & 0.099**                                                  & -0.041   & -0.04                                                & -0.101**                                              & 0.087*                                             & -0.087*                                             & 0.127**                                                & -0.02         & 0.125**                                             & -0.125**                                           \\
5466 & Crypto   & -0.014                                                   & -0.006                                                   & -0.06    & -0.048                                               & -0.121**                                              & 0.115**                                            & -0.115**                                            & -0.038                                                 & -0.025        & 0.1**                                               & -0.1**                                             \\
5470 & Pop      & -0.03                                                    & -0.025                                                   & -0.092** & -0.087*                                              & -0.164**                                              & 0.131**                                            & -0.131**                                            & -0.056                                                 & -0.069*       & 0.063                                               & -0.063                                             \\
5476 & NFTs     & -0.114**                                                 & 0.193**                                                  & -0.14**  & -0.124**                                             & -0.292**                                              & 0.308**                                            & -0.308**                                            & 0.028                                                  & -0.152**      & -0.019                                              & 0.019                                              \\
5478 & Science  & -0.099**                                                 & 0.034                                                    & -0.111** & -0.113**                                             & -0.121**                                              & 0.097**                                            & -0.097**                                            & 0.029                                                  & -0.107**      & 0.031                                               & -0.031                                             \\
5481 & Politics & 0.023                                                    & 0.055                                                    & 0.009    & 0.03                                                 & -0.134**                                              & 0.251**                                            & -0.251**                                            & -0.096**                                               & -0.017        & 0.261**                                             & -0.261**                                           \\
5487 & Tennis   & -0.012                                                   & -0.036                                                   & -0.041   & -0.035                                               & -0.103**                                              & 0.115**                                            & -0.115**                                            & -0.004                                                 & -0.091**      & 0.059                                               & -0.059                                             \\
5546 & AI       & -0.011                                                   & 0.311**                                                  & -0.081*  & -0.075*                                              & 0.037                                                 & -0.037                                             & 0.037                                               & -0.059                                                 & -0.098**      & 0.143**                                             & -0.143**                                           \\ \bottomrule
\end{tabular}
}
\caption{Correlation Analysis Results: Event-level Features} 
\label{tab:app-corr-event}
\end{table*}

\textbf{Risk preference features}: We introduced several metrics to quantify users' risk preferences. First, we calculated the average price difference ($F_{\text{avgPriceDiff}}$) between each trade's price and the midpoint (0.5), which measures the degree to which users are willing to deviate from the neutral or uncertain price. We also computed the volume-weighted average price difference ($F_{\text{volWeightedAvgPriceDiff}}$) to account for the relative importance of each trade based on its volume. Moreover, we assessed users' concentration in different events and markets using the Herfindahl-Hirschman Index (HHI)~\cite{rhoades1993herfindahl}, where the market concentration ($F_{\text{marketConcentration}}$) and event concentration ($F_{\text{eventConcentration}}$) provide insights into user addresses' risk diversification strategies and their focus on specific segments:
\begin{equation}
F_{\text{avgPriceDiff}} = \frac{\sum_{i=1}^{n} |p_i - 0.5|}{n}
\end{equation}
\begin{equation}
F_{\text{volWeightedAvgPriceDiff}} = \frac{\sum_{i=1}^{n} |p_i - 0.5| \cdot \frac{v_i}{V}}{\sum_{i=1}^{n} \frac{v_i}{V}}
\end{equation}
\begin{equation}
F_{\text{marketConcentration}} = \frac{|\{C_i | i \in T\}|}{|T|}
\end{equation}
\begin{equation}
F_{\text{eventConcentration}} = \frac{|\{EC_i | i \in T\}|}{|T|}
\end{equation}
, where $p_i$ is the price of trade $i$, $n$ is the total number of trades, $v_i$ is the volume of trade $i$, $V$ is the total volume traded, $C_i$ is the market category of trade $i$, $EC_i$ is the event category of trade $i$, and $T$ is the set of all trades.

\textbf{Combination features}: In addition to the features above, we derived several combination features to capture more subtle aspects of user address behavior: average trade amount ($F_{\text{avgTradeAmount}}$): The ratio of total volume traded to the total number of activities, reflecting the average size of a user's trades; average amount per market ($F_{\text{avgAmountPerMarket}}$) and average amount per event ($F_{\text{avgAmountPerEvent}}$): The ratio of total volume traded to the number of markets or events traded, indicating the average investment; accuracy-frequency product ($F_{\text{accFreqProduct}}$): the multiplication of the user's success rate by the total number of bets placed within the political event category (id 5481). This metric aims to encapsulate the interplay between predictive accuracy and the level of engagement within the political betting sphere. This approach aligns with recent research indicating that expertise in a specific area often correlates with the quality and quantity of engagement in related tasks~\cite{rich2010job,bakker2012work}:
\begin{equation}
F_{\text{avgTradeAmount}} = \frac{V}{A}
\end{equation}
\begin{equation}
F_{\text{avgAmountPerMarket}} = \frac{V}{M}
\end{equation}
\begin{equation}
F_{\text{avgAmountPerEvent}} = \frac{V}{E}
\end{equation}
\begin{equation}
F_{\text{accFreqProduct}} = F_{\text{successRate}} \cdot (F_{\text{buyFrequency}} + F_{\text{sellFrequency}})
\end{equation}
, where $V$ is the total volume traded, $A$ is the total number of activities, $M$ is the number of markets traded, $E$ is the number of events traded.

\section{Correlation Analysis Results}\label{app:corr}
Tables \ref{tab:app-corr-user}, \ref{tab:app-corr-event}, and \ref{tab:app-corr-market} present the correlation between features and PBLS at the user, event, and market levels, respectively. The Spearman coefficients are reported, with asterisks indicating the level of statistical significance.

\section{Panel Data Analysis Results}\label{app:panel}
In these output below, ``const'' refers to the intercept ($\theta_0$), while ``PBLS'', ``democratic\_price'' (or ``republican\_price''), and ``PBLS\_dem\_price\_interaction'' (or ``PBLS\_rep\_price\_inter\\action'') correspond to $\theta_1$, $\theta_2$, and $\theta_3$, respectively. Tables~\ref{tab:pan_total_dem} and~\ref{tab:pan_total_rep} present results for the entire duration, considering Dem and Rep, respectively. Tables \ref{tab:panel_bef_dem} and \ref{tab:panel_bef_rep}, tables \ref{tab:panel_aft_dem} and \ref{tab:panel_aft_rep} show the results before and after the election night, respectively. 

All the F-statistics are highly significant (p < 0.001), indicating that the predictors collectively impact the holding size. While R-squared may seem low, they are common in panel data settings where substantial unexplained heterogeneity exists across individuals and time periods~\cite{newman2000discussion}.

Our primary purpose is to investigate the impact of political leaning and prices on holding sizes, rather than to maximize the model fit. The low R-squared values do not necessarily undermine the significance or relevance of our findings. The consistent significance of the predictors across different periods suggests that political leaning and prices play a meaningful role in shaping investment decisions~\cite{brown1999use}. Future research could explore additional predictors or alternative methods to improve the explanation.

\begin{table*}[htbp]
\centering
\renewcommand\arraystretch{1}
\resizebox{0.9\linewidth}{!}{ 
\begin{tabular}{cc|ccccccccccc}
\toprule
\textbf{ID}   & \textbf{Name}                                                        & \begin{tabular}[c]{@{}c@{}}\textbf{category}\\ \textbf{Count}\end{tabular} & \begin{tabular}[c]{@{}c@{}}\textbf{category}\\ \textbf{Ratio}\end{tabular} & \textbf{Amount}   & \begin{tabular}[c]{@{}c@{}}\textbf{buy}\\ \textbf{Amount}\end{tabular} & \begin{tabular}[c]{@{}c@{}}\textbf{sell}\\ \textbf{Amount}\end{tabular} & \begin{tabular}[c]{@{}c@{}}\textbf{buy}\\ \textbf{Freq}\end{tabular} & \begin{tabular}[c]{@{}c@{}}\textbf{sell}\\ \textbf{Freq}\end{tabular} & \begin{tabular}[c]{@{}c@{}}\textbf{success}\\ \textbf{Rate}\end{tabular} & \textbf{profitability} & \begin{tabular}[c]{@{}c@{}}\textbf{yes}\\ \textbf{Ratio}\end{tabular} & \begin{tabular}[c]{@{}c@{}}\textbf{no}\\ \textbf{Ratio}\end{tabular} \\ \midrule
5456 & Business                                                    & -0.058                                                   & 0.143**                                                  & 0.086*   & 0.086*                                               & N/A                                                   & N/A                                                & N/A                                                 & 1**                                                    & 1**           & 0.778**                                             & -0.778**                                           \\
5457 & Inflation                                                   & -0.238**                                                 & 0.178**                                                  & -0.095** & -0.081*                                              & -0.34**                                               & 0.303**                                            & -0.303**                                            & 0.23**                                                 & 0.092**       & -0.069*                                             & 0.069*                                             \\
5458 & \begin{tabular}[c]{@{}c@{}}Interest \\ rates\end{tabular}   & 0.071*                                                   & 0.12**                                                   & 0.07*    & 0.082*                                               & -0.069*                                               & 0.097**                                            & -0.097**                                            & -0.018                                                 & -0.151**      & 0.032                                               & -0.032                                             \\
5459 & \begin{tabular}[c]{@{}c@{}}Commodity \\ prices\end{tabular} & -0.166**                                                 & 0.204**                                                  & -0.162** & -0.163**                                             & -0.204**                                              & 0.207**                                            & -0.207**                                            & 0.422**                                                & -0.092**      & -0.05                                               & 0.05                                               \\
5460 & Forex                                                       & -0.155**                                                 & 0.234**                                                  & -0.088** & -0.103**                                             & -0.217**                                              & 0.19**                                             & -0.19**                                             & 0.1**                                                  & 0.106**       & 0.219**                                             & -0.219**                                           \\
5461 & Unemploy                                                    & -0.254**                                                 & -0.13**                                                  & 0.063    & 0.063                                                & N/A                                                   & N/A                                                & N/A                                                 & 0.658**                                                & 0.424**       & 0.147**                                             & -0.147**                                           \\
5462 & Tech                                                        & -0.119**                                                 & 0.05                                                     & -0.178** & -0.181**                                             & -0.144**                                              & 0.105**                                            & -0.105**                                            & 0.027                                                  & 0.19**        & 0.138**                                             & -0.138**                                           \\
5463 & Coronavirus                                                 & 0.026                                                    & -0.023                                                   & -0.049   & -0.05                                                & -0.07*                                                & 0.037                                              & -0.037                                              & 0.464**                                                & 0.6**         & 0.059                                               & -0.059                                             \\
5464 & Cases                                                       & 0.003                                                    & 0.221**                                                  & 0.06     & 0.06                                                 & -0.167**                                              & 0.167**                                            & -0.167**                                            & 0.123**                                                & 0.153**       & 0.08*                                               & -0.08*                                             \\
5465 & Vaccinations                                                & -0.035                                                   & 0.105**                                                  & 0.049    & 0.049                                                & -0.047                                                & 0.047                                              & -0.047                                              & -0.021                                                 & -0.025        & 0.003                                               & -0.003                                             \\
5466 & Crypto                                                      & -0.092**                                                 & 0.187**                                                  & -0.134** & -0.131**                                             & 0.165**                                               & -0.144**                                           & 0.144**                                             & 0.031                                                  & -0.268**      & -0.021                                              & 0.021                                              \\
5467 & Prices                                                      & -0.033                                                   & -0.014                                                   & -0.058   & -0.045                                               & -0.166**                                              & 0.157**                                            & -0.157**                                            & -0.014                                                 & 0.039         & 0.074*                                              & -0.074*                                            \\
5468 & Airdrops                                                    & -0.059                                                   & -0.002                                                   & -0.023   & -0.023                                               & -0.102**                                              & 0.099**                                            & -0.099**                                            & 0.203**                                                & -0.022        & 0.17**                                              & -0.17**                                            \\
5469 & Stablecoins                                                 & -0.134**                                                 & 0.198**                                                  & -0.101** & -0.113**                                             & -0.095**                                              & 0.122**                                            & -0.122**                                            & 0.072*                                                 & 0.349**       & 0.079*                                              & -0.079*                                            \\
5470 & Pop                                                         & 0.425**                                                  & 0.349**                                                  & 0.813**  & 0.813**                                              & N/A                                                   & N/A                                                & N/A                                                 & -0.735**                                               & -0.167**      & 0.184**                                             & -0.184**                                           \\
5471 & Art                                                         & -0.118**                                                 & 0.12**                                                   & -0.147** & -0.145**                                             & -0.02                                                 & -0.01                                              & 0.01                                                & 0.123**                                                & 0.187**       & 0.245**                                             & -0.245**                                           \\
5472 & Celebrities                                                 & -0.097**                                                 & -0.047                                                   & -0.126** & -0.12**                                              & -0.197**                                              & 0.146**                                            & -0.146**                                            & 0.026                                                  & 0.073*        & 0.01                                                & -0.01                                              \\
5473 & \begin{tabular}[c]{@{}c@{}}Film\\ TV\end{tabular}           & -0.075*                                                  & 0.027                                                    & -0.114** & -0.108**                                             & -0.152**                                              & 0.152**                                            & -0.152**                                            & -0.038                                                 & -0.017        & -0.002                                              & 0.002                                              \\
5474 & Music                                                       & -0.129**                                                 & -0.033                                                   & -0.175** & -0.174**                                             & -0.207**                                              & 0.195**                                            & -0.195**                                            & 0.132**                                                & 0.324**       & 0.205**                                             & -0.205**                                           \\
5475 & Twitter                                                     & -0.05                                                    & 0.095**                                                  & -0.07*   & -0.072*                                              & -0.003                                                & 0.025                                              & -0.025                                              & -0.091**                                               & -0.026        & -0.003                                              & 0.003                                              \\
5476 & NFTs                                                        & -0.31**                                                  & 0.188**                                                  & -0.294** & -0.234**                                             & -0.205**                                              & 0.257**                                            & -0.257**                                            & -0.029                                                 & -0.32**       & 0.233**                                             & -0.233**                                           \\
5477 & \begin{tabular}[c]{@{}c@{}}Floor \\ prices\end{tabular}     & -0.024                                                   & 0.208**                                                  & -0.083*  & -0.068*                                              & -0.242**                                              & 0.252**                                            & -0.252**                                            & 0.034                                                  & 0.386**       & -0.097**                                            & 0.097**                                            \\
5478 & Science                                                     & N/A                                                      & 0.4**                                                    & -0.6**   & -0.6**                                               & N/A                                                   & N/A                                                & N/A                                                 & N/A                                                    & N/A           & 0.577**                                             & -0.577**                                           \\
5479 & Climate                                                     & -0.122**                                                 & 0.079*                                                   & -0.169** & -0.179**                                             & -0.119**                                              & 0.117**                                            & -0.117**                                            & 0.062                                                  & -0.136**      & -0.025                                              & 0.025                                              \\
5480 & Space                                                       & -0.084*                                                  & 0.116**                                                  & -0.136** & -0.151**                                             & -0.081*                                               & 0.049                                              & -0.049                                              & 0.064                                                  & 0.14**        & 0.046                                               & -0.046                                             \\
5481 & Politics                                                    & -0.224**                                                 & 0.272**                                                  & -0.303** & -0.291**                                             & -0.289**                                              & 0.275**                                            & -0.275**                                            & -0.085*                                                & -0.165**      & 0.112**                                             & -0.112**                                           \\
5482 & Elections                                                   & 0.017                                                    & 0.029                                                    & -0.011   & 0.012                                                & -0.163**                                              & 0.278**                                            & -0.278**                                            & -0.031                                                 & -0.061        & 0.299**                                             & -0.299**                                           \\
5483 & \begin{tabular}[c]{@{}c@{}}Global \\ Politics\end{tabular}  & 0.001                                                    & -0.023                                                   & -0.059   & -0.054                                               & -0.09**                                               & 0.089**                                            & -0.089**                                            & -0.067                                                 & -0.002        & 0.036                                               & -0.036                                             \\
5484 & Polls                                                       & -0.093**                                                 & 0.225**                                                  & -0.129** & -0.132**                                             & -0.093**                                              & 0.028                                              & -0.028                                              & 0                                                      & 0.135**       & 0.065                                               & -0.065                                             \\
5485 & Bills                                                       & -0.087*                                                  & 0.186**                                                  & -0.035   & -0.027                                               & -0.133**                                              & 0.113**                                            & -0.113**                                            & 0.083*                                                 & 0.094**       & -0.009                                              & 0.009                                              \\
5486 & Trump                                                       & -0.074*                                                  & 0.069*                                                   & -0.099** & -0.101**                                             & -0.062                                                & 0.028                                              & -0.028                                              & -0.02                                                  & 0.111**       & 0.022                                               & -0.022                                             \\
5487 & Sports                                                      & -0.296**                                                 & 0.061                                                    & -0.121** & -0.126**                                             & -0.32**                                               & 0.326**                                            & -0.326**                                            & 0                                                      & -0.866**      & -0.048                                              & 0.048                                              \\
5488 & Basketball                                                  & -0.105**                                                 & 0.096**                                                  & -0.098** & -0.095**                                             & -0.014                                                & -0.003                                             & 0.003                                               & 0.004                                                  & 0.148**       & 0.02                                                & -0.02                                              \\
5489 & Golf                                                        & -0.033                                                   & 0.114**                                                  & -0.002   & -0.018                                               & -0.089**                                              & 0.1**                                              & -0.1**                                              & 0.54**                                                 & 0.378**       & -0.073*                                             & 0.073*                                             \\
5490 & Chess                                                       & 0.008                                                    & 0.293**                                                  & -0.186** & -0.191**                                             & -0.029                                                & 0.042                                              & -0.042                                              & 0.046                                                  & 0.04          & 0.116**                                             & -0.116**                                           \\
5491 & Boxing                                                      & -0.197**                                                 & 0.003                                                    & -0.163** & -0.167**                                             & -0.176**                                              & 0.151**                                            & -0.151**                                            & 0.086*                                                 & 0.121**       & 0.154**                                             & -0.154**                                           \\
5492 & Football                                                    & -0.118**                                                 & -0.022                                                   & -0.148** & -0.141**                                             & -0.163**                                              & 0.139**                                            & -0.139**                                            & -0.033                                                 & 0.076*        & 0.233**                                             & -0.233**                                           \\
5493 & Esports                                                     & -0.182**                                                 & 0.263**                                                  & -0.293** & -0.306**                                             & -0.056                                                & -0.057                                             & 0.057                                               & 0.28**                                                 & -0.132**      & 0.011                                               & -0.011                                             \\
5494 & Soccer                                                      & -0.024                                                   & 0.124**                                                  & -0.114** & -0.104**                                             & -0.132**                                              & 0.136**                                            & -0.136**                                            & -0.113**                                               & 0.019         & 0.134**                                             & -0.134**                                           \\
5495 & Racing                                                      & 0.192**                                                  & 0.363**                                                  & -0.032   & -0.037                                               & 0.239**                                               & -0.206**                                           & 0.206**                                             & 0.004                                                  & 0.104**       & -0.034                                              & 0.034                                              \\
5496 & Baseball                                                    & -0.074*                                                  & 0.166**                                                  & -0.18**  & -0.19**                                              & -0.19**                                               & 0.198**                                            & -0.198**                                            & 0.169**                                                & -0.124**      & 0.369**                                             & -0.369**                                           \\
5497 & Exploits                                                    & -0.191**                                                 & 0.012                                                    & -0.234** & -0.226**                                             & -0.11**                                               & 0.05                                               & -0.05                                               & 0.169**                                                & -0.254**      & 0.356**                                             & -0.356**                                           \\
5498 & Hockey                                                      & -0.397**                                                 & -0.024                                                   & -0.259** & -0.276**                                             & -0.11**                                               & 0.088**                                            & -0.088**                                            & 0.555**                                                & -0.021        & -0.006                                              & 0.006                                              \\
5499 & Travel                                                      & -0.26**                                                  & 0.059                                                    & -0.175** & -0.156**                                             & -0.135**                                              & 0.183**                                            & -0.183**                                            & 0.208**                                                & -0.236**      & 0.174**                                             & -0.174**                                           \\
5500 & Academia                                                    & -0.089**                                                 & 0.093**                                                  & -0.051   & -0.053                                               & -0.086*                                               & 0.067*                                             & -0.067*                                             & -0.027                                                 & 0.278**       & 0.081*                                              & -0.081*                                            \\
5501 & Cricket                                                     & -0.366**                                                 & 0.147**                                                  & -0.379** & -0.379**                                             & -0.534**                                              & 0.466**                                            & -0.466**                                            & 0.11**                                                 & 0.45**        & 0.327**                                             & -0.327**                                           \\
5502 & Tennis                                                      & -0.138**                                                 & 0.11**                                                   & -0.223** & -0.157**                                             & -0.237**                                              & 0.231**                                            & -0.231**                                            & 0.185**                                                & 0.006         & 0.123**                                             & -0.123**                                           \\
5503 & Billionaires                                                & -0.077*                                                  & 0.13**                                                   & -0.125** & -0.12**                                              & -0.006                                                & -0.037                                             & 0.037                                               & 0.069*                                                 & -0.162**      & -0.076*                                             & 0.076*                                             \\
5504 & Marketcaps                                                  & -0.27**                                                  & 0.169**                                                  & -0.141** & -0.103**                                             & -0.299**                                              & 0.354**                                            & -0.354**                                            & 0.138**                                                & 0.123**       & 0.069*                                              & -0.069*                                            \\
5505 & Olympics                                                    & -0.042                                                   & 0.189**                                                  & 0.034    & 0.034                                                & N/A                                                   & N/A                                                & N/A                                                 & 0.11**                                                 & -0.101**      & 0.087*                                              & -0.087*                                            \\
5506 & Poker                                                       & -0.261**                                                 & 0.284**                                                  & -0.152** & -0.152**                                             & 0.205**                                               & -0.205**                                           & 0.205**                                             & 0.141**                                                & 0.333**       & 0.007                                               & -0.007                                             \\
5507 & Variants                                                    & -0.433**                                                 & 0.252**                                                  & -0.298** & -0.295**                                             & -0.261**                                              & 0.217**                                            & -0.217**                                            & 0.264**                                                & 0.572**       & 0.17**                                              & -0.17**                                            \\
5508 & Courts                                                      & -0.101**                                                 & 0.073*                                                   & -0.205** & -0.21**                                              & -0.1**                                                & 0.08*                                              & -0.08*                                              & 0.095**                                                & 0.086*        & 0.033                                               & -0.033                                             \\
5509 & Biden                                                       & 0.028                                                    & 0.328**                                                  & 0.005    & 0                                                    & 0.015                                                 & 0.025                                              & -0.025                                              & -0.176**                                               & 0.056         & -0.148**                                            & 0.148**                                            \\
5510 & Finance                                                     & -0.035                                                   & -0.09**                                                  & -0.01    & 0.002                                                & -0.106**                                              & 0.103**                                            & -0.103**                                            & 0.05                                                   & 0.006         & -0.005                                              & 0.005                                              \\
5511 & Biking                                                      & -0.087*                                                  & 0.555**                                                  & -0.173** & -0.155**                                             & 0.321**                                               & 0.074*                                             & -0.074*                                             & -0.192**                                               & 0.429**       & 0.138**                                             & -0.138**                                           \\
5512 & Exchanges                                                   & -0.117**                                                 & 0.011                                                    & -0.189** & -0.179**                                             & -0.249**                                              & 0.173**                                            & -0.173**                                            & 0.171**                                                & 0.284**       & 0.085*                                              & -0.085*                                            \\
5545 & US politics                                                 & 0.061                                                    & 0.07*                                                    & 0.062    & 0.062                                                & -0.021                                                & 0.024                                              & -0.024                                              & -0.09**                                                & -0.074*       & 0.07*                                               & -0.07*                                             \\
5546 & AI                                                          & 0.049                                                    & 0.291**                                                  & 0.038    & 0.041                                                & 0.039                                                 & -0.044                                             & 0.044                                               & -0.134**                                               & -0.424**      & 0.198**                                             & -0.198**                                           \\
5547 & Chat Bots                                                   & -0.074*                                                  & 0.435**                                                  & -0.126** & -0.139**                                             & 0.034                                                 & -0.037                                             & 0.037                                               & 0.005                                                  & 0.09**        & -0.001                                              & 0.001                                              \\
5548 & Friend Tech                                                 & -0.379**                                                 & 0.036                                                    & -0.438** & -0.438**                                             & -0.345**                                              & 0.312**                                            & -0.312**                                            & 0.569**                                                & -0.596**      & -0.147**                                            & 0.147**                                            \\
5549 & Middle East                                                 & -0.08*                                                   & -0.003                                                   & -0.147** & -0.144**                                             & -0.137**                                              & 0.129**                                            & -0.129**                                            & -0.003                                                 & -0.001        & 0.085*                                              & -0.085*                                            \\
5550 & Oil                                                         & -0.056                                                   & 0.193**                                                  & -0.015   & -0.034                                               & 0.055                                                 & -0.024                                             & 0.024                                               & -0.164**                                               & 0.051         & 0.059                                               & -0.059                                             \\ \bottomrule
\end{tabular}
}
\caption{Correlation Analysis Results: Market-level Features} 
\label{tab:app-corr-market}
\end{table*}

\clearpage

\begin{table}[!htb]
\centering
\renewcommand\arraystretch{0.6}
\resizebox{\linewidth}{!}{ 
\begin{tabular}{lclc}
\toprule
\textbf{Dep. Variable:}                & dem\_holding\_size & \textbf{  R-squared:         }   &      0.0079      \\
\textbf{Estimator:}                    &   RandomEffects    & \textbf{  R-squared (Between):}  &      0.0219      \\
\textbf{No. Observations:}             &       22624        & \textbf{  R-squared (Within):}   &      0.0076      \\
\textbf{Date:}                         &  Thu, Mar 21 2024  & \textbf{  R-squared (Overall):}  &      0.0320      \\
\textbf{Time:}                         &      00:30:40      & \textbf{  Log-likelihood     }   &    -2.322e+05    \\
\textbf{Cov. Estimator:}               &     Unadjusted     & \textbf{                     }   &                  \\
\textbf{}                              &                    & \textbf{  F-statistic:       }   &      60.387      \\
\textbf{Entities:}                     &        373         & \textbf{  P-value            }   &      0.0000      \\
\textbf{Avg Obs:}                      &       60.654       & \textbf{  Distribution:      }   &    F(3,22620)    \\
\textbf{Min Obs:}                      &       1.0000       & \textbf{                     }   &                  \\
\textbf{Max Obs:}                      &       306.00       & \textbf{  F-statistic (robust):} &      59.891      \\
\textbf{}                              &                    & \textbf{  P-value            }   &      0.0000      \\
\textbf{Time periods:}                 &        306         & \textbf{  Distribution:      }   &    F(3,22620)    \\
\textbf{Avg Obs:}                      &       73.935       & \textbf{                     }   &                  \\
\textbf{Min Obs:}                      &       4.0000       & \textbf{                     }   &                  \\
\textbf{Max Obs:}                      &       373.00       & \textbf{                     }   &                  \\
\textbf{}                              &                    & \textbf{                     }   &                  \\
\bottomrule
\end{tabular}
}
\resizebox{\linewidth}{!}{
\begin{tabular}{lcccccc}
                                       & \textbf{Parameter} & \textbf{Std. Err.} & \textbf{T-stat} & \textbf{P-value} & \textbf{Lower CI} & \textbf{Upper CI}  \\
\midrule
\textbf{const}                         &       964.79       &       1130.4       &      0.8535     &      0.3934      &      -1250.8      &       3180.4       \\
\textbf{PBLS}                          &       1542.6       &       1214.0       &      1.2707     &      0.2038      &      -836.89      &       3922.1       \\
\textbf{democratic\_price}             &       587.85       &       294.62       &      1.9953     &      0.0460      &       10.370      &       1165.3       \\
\textbf{PBLS\_dem\_price\_interaction} &       3452.7       &       317.20       &      10.885     &      0.0000      &       2831.0      &       4074.4       \\
\bottomrule
\end{tabular}
}

\caption{Panel Regression Results for Democratic Holding Size (Total Duration)}
\label{tab:pan_total_dem}
\end{table} 

\begin{table}[!htb] 
\centering
\renewcommand\arraystretch{0.6}
\resizebox{\linewidth}{!}{ 
\begin{tabular}{lclc}
\toprule
\textbf{Dep. Variable:}                & rep\_holding\_size & \textbf{  R-squared:         }   &      0.0035      \\
\textbf{Estimator:}                    &   RandomEffects    & \textbf{  R-squared (Between):}  &      0.0078      \\
\textbf{No. Observations:}             &       22624        & \textbf{  R-squared (Within):}   &      0.0028      \\
\textbf{Date:}                         &  Thu, Mar 21 2024  & \textbf{  R-squared (Overall):}  &     0.0019      \\
\textbf{Time:}                         &      00:30:41      & \textbf{  Log-likelihood     }   &    -2.121e+05    \\
\textbf{Cov. Estimator:}               &     Unadjusted     & \textbf{                     }   &                  \\
\textbf{}                              &                    & \textbf{  F-statistic:       }   &      26.642      \\
\textbf{Entities:}                     &        373         & \textbf{  P-value            }   &      0.0000      \\
\textbf{Avg Obs:}                      &       60.654       & \textbf{  Distribution:      }   &    F(3,22620)    \\
\textbf{Min Obs:}                      &       1.0000       & \textbf{                     }   &                  \\
\textbf{Max Obs:}                      &       306.00       & \textbf{  F-statistic (robust):} &      21.884      \\
\textbf{}                              &                    & \textbf{  P-value            }   &      0.0000      \\
\textbf{Time periods:}                 &        306         & \textbf{  Distribution:      }   &    F(3,22620)    \\
\textbf{Avg Obs:}                      &       73.935       & \textbf{                     }   &                  \\
\textbf{Min Obs:}                      &       4.0000       & \textbf{                     }   &                  \\
\textbf{Max Obs:}                      &       373.00       & \textbf{                     }   &                  \\
\textbf{}                              &                    & \textbf{                     }   &                  \\
\bottomrule
\end{tabular}
}
\resizebox{\linewidth}{!}{
\begin{tabular}{lcccccc}
                                       & \textbf{Parameter} & \textbf{Std. Err.} & \textbf{T-stat} & \textbf{P-value} & \textbf{Lower CI} & \textbf{Upper CI}  \\
\midrule
\textbf{const}                         &       2394.5       &       369.06       &      6.4881     &      0.0000      &       1671.1      &       3117.9       \\
\textbf{PBLS}                          &      -659.56       &       396.41       &     -1.6638     &      0.0962      &      -1436.6      &       117.44       \\
\textbf{republican\_price}             &      -953.42       &       121.03       &     -7.8774     &      0.0000      &      -1190.6      &      -716.19       \\
\textbf{PBLS\_rep\_price\_interaction} &       279.45       &       130.31       &      2.1445     &      0.0320      &       24.034      &       534.88       \\
\bottomrule
\end{tabular}
}
\caption{Panel Regression Results for Republican Holding Size (Total Duration)}
\label{tab:pan_total_rep}
\end{table}

\begin{table}[!htb]
\centering
\renewcommand\arraystretch{0.6}
\resizebox{\linewidth}{!}{ 
\begin{tabular}{lclc}
\toprule
\textbf{Dep. Variable:}                & dem\_holding\_size & \textbf{  R-squared:         }   &      0.0013      \\
\textbf{Estimator:}                    &   RandomEffects    & \textbf{  R-squared (Between):}  &      0.0319      \\
\textbf{No. Observations:}             &       19897        & \textbf{  R-squared (Within):}   &      0.0008      \\
\textbf{Date:}                         &  Thu, Mar 21 2024  & \textbf{  R-squared (Overall):}  &      0.0320      \\
\textbf{Time:}                         &      00:32:50      & \textbf{  Log-likelihood     }   &    -2.008e+05    \\
\textbf{Cov. Estimator:}               &     Unadjusted     & \textbf{                     }   &                  \\
\textbf{}                              &                    & \textbf{  F-statistic:       }   &      8.7418      \\
\textbf{Entities:}                     &        273         & \textbf{  P-value            }   &      0.0000      \\
\textbf{Avg Obs:}                      &       72.883       & \textbf{  Distribution:      }   &    F(3,19893)    \\
\textbf{Min Obs:}                      &       1.0000       & \textbf{                     }   &                  \\
\textbf{Max Obs:}                      &       298.00       & \textbf{  F-statistic (robust):} &      8.2685      \\
\textbf{}                              &                    & \textbf{  P-value            }   &      0.0000      \\
\textbf{Time periods:}                 &        298         & \textbf{  Distribution:      }   &    F(3,19893)    \\
\textbf{Avg Obs:}                      &       66.768       & \textbf{                     }   &                  \\
\textbf{Min Obs:}                      &       4.0000       & \textbf{                     }   &                  \\
\textbf{Max Obs:}                      &       273.00       & \textbf{                     }   &                  \\
\textbf{}                              &                    & \textbf{                     }   &                  \\
\bottomrule
\end{tabular}

}
\resizebox{\linewidth}{!}{
\begin{tabular}{lcccccc}
                                       & \textbf{Parameter} & \textbf{Std. Err.} & \textbf{T-stat} & \textbf{P-value} & \textbf{Lower CI} & \textbf{Upper CI}  \\
\midrule
\textbf{const}                         &       1100.6       &       1297.1       &      0.8485     &      0.3962      &      -1441.9      &       3643.1       \\
\textbf{PBLS}                          &       3464.3       &       1396.8       &      2.4802     &      0.0131      &       726.51      &       6202.0       \\
\textbf{democratic\_price}             &      -42.489       &       339.51       &     -0.1251     &      0.9004      &      -707.96      &       622.98       \\
\textbf{PBLS\_dem\_price\_interaction} &       1360.5       &       368.06       &      3.6964     &      0.0002      &       639.08      &       2081.9       \\
\bottomrule
\end{tabular}
}
\caption{Panel Regression for Democratic Holding Size (Before Election Night)}
\label{tab:panel_bef_dem}
\end{table}

\begin{table}[!htb]
\centering
\renewcommand\arraystretch{0.6}
\resizebox{\linewidth}{!}{ 
\begin{tabular}{lclc}
\toprule
\textbf{Dep. Variable:}                & rep\_holding\_size & \textbf{  R-squared:         }   &      0.0068      \\
\textbf{Estimator:}                    &   RandomEffects    & \textbf{  R-squared (Between):}  &      0.0037      \\
\textbf{No. Observations:}             &       19897        & \textbf{  R-squared (Within):}   &      0.0066      \\
\textbf{Date:}                         &  Thu, Mar 21 2024  & \textbf{  R-squared (Overall):}  &     0.0052      \\
\textbf{Time:}                         &      00:32:51      & \textbf{  Log-likelihood     }   &    -1.803e+05    \\
\textbf{Cov. Estimator:}               &     Unadjusted     & \textbf{                     }   &                  \\
\textbf{}                              &                    & \textbf{  F-statistic:       }   &      45.418      \\
\textbf{Entities:}                     &        273         & \textbf{  P-value            }   &      0.0000      \\
\textbf{Avg Obs:}                      &       72.883       & \textbf{  Distribution:      }   &    F(3,19893)    \\
\textbf{Min Obs:}                      &       1.0000       & \textbf{                     }   &                  \\
\textbf{Max Obs:}                      &       298.00       & \textbf{  F-statistic (robust):} &      44.101      \\
\textbf{}                              &                    & \textbf{  P-value            }   &      0.0000      \\
\textbf{Time periods:}                 &        298         & \textbf{  Distribution:      }   &    F(3,19893)    \\
\textbf{Avg Obs:}                      &       66.768       & \textbf{                     }   &                  \\
\textbf{Min Obs:}                      &       4.0000       & \textbf{                     }   &                  \\
\textbf{Max Obs:}                      &       273.00       & \textbf{                     }   &                  \\
\textbf{}                              &                    & \textbf{                     }   &                  \\
\bottomrule
\end{tabular}

}
\resizebox{\linewidth}{!}{
\begin{tabular}{lcccccc}
                                       & \textbf{Parameter} & \textbf{Std. Err.} & \textbf{T-stat} & \textbf{P-value} & \textbf{Lower CI} & \textbf{Upper CI}  \\
\midrule
\textbf{const}                         &       1290.5       &       381.08       &      3.3865     &      0.0007      &       543.59      &       2037.5       \\
\textbf{PBLS}                          &      -1186.6       &       410.78       &     -2.8886     &      0.0039      &      -1991.7      &      -381.43       \\
\textbf{republican\_price}             &       399.28       &       121.57       &      3.2844     &      0.0010      &       161.00      &       637.56       \\
\textbf{PBLS\_rep\_price\_interaction} &       1116.3       &       131.79       &      8.4701     &      0.0000      &       857.96      &       1374.6       \\
\bottomrule
\end{tabular}
}
\caption{Panel Regression for Republican Holding Size (Before Election Night)}
\label{tab:panel_bef_rep}
\end{table}

\begin{table}[!htb]
\centering
\renewcommand\arraystretch{0.6}
\resizebox{\linewidth}{!}{ 
\begin{tabular}{lclc}
\toprule
\textbf{Dep. Variable:}                & dem\_holding\_size & \textbf{  R-squared:         }   &      0.0411      \\
\textbf{Estimator:}                    &   RandomEffects    & \textbf{  R-squared (Between):}  &      0.0384      \\
\textbf{No. Observations:}             &        2727        & \textbf{  R-squared (Within):}   &      0.0419      \\
\textbf{Date:}                         &  Thu, Mar 21 2024  & \textbf{  R-squared (Overall):}  &      0.0413      \\
\textbf{Time:}                         &      00:34:16      & \textbf{  Log-likelihood     }   &    -2.531e+04    \\
\textbf{Cov. Estimator:}               &     Unadjusted     & \textbf{                     }   &                  \\
\textbf{}                              &                    & \textbf{  F-statistic:       }   &      38.934      \\
\textbf{Entities:}                     &        373         & \textbf{  P-value            }   &      0.0000      \\
\textbf{Avg Obs:}                      &       7.3110       & \textbf{  Distribution:      }   &    F(3,2723)     \\
\textbf{Min Obs:}                      &       1.0000       & \textbf{                     }   &                  \\
\textbf{Max Obs:}                      &       8.0000       & \textbf{  F-statistic (robust):} &      39.174      \\
\textbf{}                              &                    & \textbf{  P-value            }   &      0.0000      \\
\textbf{Time periods:}                 &         8          & \textbf{  Distribution:      }   &    F(3,2723)     \\
\textbf{Avg Obs:}                      &       340.88       & \textbf{                     }   &                  \\
\textbf{Min Obs:}                      &       276.00       & \textbf{                     }   &                  \\
\textbf{Max Obs:}                      &       373.00       & \textbf{                     }   &                  \\
\textbf{}                              &                    & \textbf{                     }   &                  \\
\bottomrule
\end{tabular}

}
\resizebox{\linewidth}{!}{
\begin{tabular}{lcccccc}
                                       & \textbf{Parameter} & \textbf{Std. Err.} & \textbf{T-stat} & \textbf{P-value} & \textbf{Lower CI} & \textbf{Upper CI}  \\
\midrule
\textbf{const}                         &       1569.8       &       1458.5       &      1.0763     &      0.2819      &      -1290.0      &       4429.6       \\
\textbf{PBLS}                          &       4826.1       &       1566.3       &      3.0812     &      0.0021      &       1754.8      &       7897.4       \\
\textbf{democratic\_price}             &       278.78       &       194.02       &      1.4369     &      0.1509      &      -101.66      &       659.21       \\
\textbf{PBLS\_dem\_price\_interaction} &       1777.1       &       206.81       &      8.5930     &      0.0000      &       1371.6      &       2182.6       \\
\bottomrule
\end{tabular}
}
\caption{Panel Regression for Democratic Holding Size (After Election Night)}
\label{tab:panel_aft_dem}
\end{table}

\begin{table}[!htb]
\centering
\renewcommand\arraystretch{0.6}
\resizebox{\linewidth}{!}{ 
\begin{tabular}{lclc}
\toprule
\textbf{Dep. Variable:}                & rep\_holding\_size & \textbf{  R-squared:         }   &      0.0164      \\
\textbf{Estimator:}                    &   RandomEffects    & \textbf{  R-squared (Between):}  &      0.0012      \\
\textbf{No. Observations:}             &        2727        & \textbf{  R-squared (Within):}   &      0.0184      \\
\textbf{Date:}                         &  Thu, Mar 21 2024  & \textbf{  R-squared (Overall):}  &      0.0031      \\
\textbf{Time:}                         &      00:34:16      & \textbf{  Log-likelihood     }   &    -2.667e+04    \\
\textbf{Cov. Estimator:}               &     Unadjusted     & \textbf{                     }   &                  \\
\textbf{}                              &                    & \textbf{  F-statistic:       }   &      15.139      \\
\textbf{Entities:}                     &        373         & \textbf{  P-value            }   &      0.0000      \\
\textbf{Avg Obs:}                      &       7.3110       & \textbf{  Distribution:      }   &    F(3,2723)     \\
\textbf{Min Obs:}                      &       1.0000       & \textbf{                     }   &                  \\
\textbf{Max Obs:}                      &       8.0000       & \textbf{  F-statistic (robust):} &      14.811      \\
\textbf{}                              &                    & \textbf{  P-value            }   &      0.0000      \\
\textbf{Time periods:}                 &         8          & \textbf{  Distribution:      }   &    F(3,2723)     \\
\textbf{Avg Obs:}                      &       340.88       & \textbf{                     }   &                  \\
\textbf{Min Obs:}                      &       276.00       & \textbf{                     }   &                  \\
\textbf{Max Obs:}                      &       373.00       & \textbf{                     }   &                  \\
\textbf{}                              &                    & \textbf{                     }   &                  \\
\bottomrule
\end{tabular}
}
\resizebox{\linewidth}{!}{
\begin{tabular}{lcccccc}
                                       & \textbf{Parameter} & \textbf{Std. Err.} & \textbf{T-stat} & \textbf{P-value} & \textbf{Lower CI} & \textbf{Upper CI}  \\
\midrule
\textbf{const}                         &       3142.7       &       519.64       &      6.0478     &      0.0000      &       2123.8      &       4161.6       \\
\textbf{PBLS}                          &      -266.38       &       558.01       &     -0.4774     &      0.6331      &      -1360.5      &       827.78       \\
\textbf{republican\_price}             &      -1702.5       &       319.27       &     -5.3323     &      0.0000      &      -2328.5      &      -1076.4       \\
\textbf{PBLS\_rep\_price\_interaction} &      -502.22       &       340.40       &     -1.4754     &      0.1402      &      -1169.7      &       165.24       \\
\bottomrule
\end{tabular}
}
\caption{Panel Regression for Republican Holding Size (After Election Night)}
\label{tab:panel_aft_rep}
\end{table}

\end{document}